# Defect Spectroscopy and Non-Ionizing Energy Loss Analysis of Proton and Electron Irradiated p-type GaAs Solar Cells

C. Pellegrino,[1,2,a)] A. Gagliardi,[2,b)] and C. G. Zimmermann[1,c)]

[1)] *Airbus Defence and Space GmbH, 82024 Taufkirchen, Germany*

[2)] *Technical University of Munich, 80333 Munich, Germany*

Admittance spectroscopy combined with non-ionizing energy loss (NIEL) analysis is shown to be a powerful tool for analyzing solar cell radiation degradation, not relying on the change of macroscopic cell parameters. GaAs component cells, representative of the middle sub-cell in $Ga_{0.5}In_{0.5}P$/GaAs/Ge solar cells, were irradiated with protons and electrons in the 0.5 – 3 MeV energy range. Four irradiation-induced defects are identified in the p-type base layer. The nature of each defect is assessed by analyzing the dependence of its introduction rate on the NIEL deposited by electrons in the semiconductor. The expected linear relationship is only achieved if a unique threshold energy $E_d$ is ascribed to each defect, which ranges from 9 to 38 eV. An electron NIEL with $E_d = 21$ eV, customarily used for GaAs-based solar cell degradation analysis, is an approximation of the relative abundance of these four defects. The 21 eV value is thus a GaAs material-specific parameter, independent of the electrical device design. In addition, the type and energy of the incident particle is correlated with the relative abundance of high $E_d$ defects. The impact of each defect on the macroscopic electrical parameters of the cell, namely the open-circuit voltage $V_{OC}$, the short-circuit current density $J_{SC}$ and the recombination current density $J_{02}$, is assessed with the help of a Pearson analysis. The different effectiveness of electron and proton irradiation on parameters dominated by recombination in the depleted region, such as $V_{OC}$ or $J_{02}$, is attributed in part to the influence of the particle recoil spectra on the defect capture cross-section.

[a)] Electronic mail: carmine.pellegrino@tum.de

[b)] Electronic mail: alessio.gagliardi@tum.de

[c)] Electronic mail: claus.zimmermann@airbus.com

## I. INTRODUCTION

Solar cells operated in space are subjected to proton and electron irradiation in the MeV energy range, which generates displacements in the lattice via elastic collisions with crystal atoms. In $Ga_{0.5}In_{0.5}P$/GaAs/Ge triple-junction solar cells, the GaAs sub-cell is electrically degraded most by this non-ionizing damage and is thus the most interesting one to analyze. The number of displacements generated in GaAs by a particle spectrum characterized by a differential fluence $\phi(\xi)$ can be computed via the displacement damage dose $\mathrm{DDD} = \int_{\xi} \mathrm{NIEL}(\xi)\phi(\xi)d\xi$, where $\xi$ denotes the particle energy. The non-ionizing energy loss (NIEL) quantifies the amount of non-ionizing damage that the primary particle and all its recoils deposit in the lattice in form of displacements. The DDD is commonly used to model the degradation of the solar cell parameters following irradiation with protons and electrons[1].

An important physical parameter for the NIEL determination is the threshold energy for displacement $E_{\mathrm{d}_0}$, which corresponds to the minimum energy required for a recoil atom to displace a Ga or As atom from its lattice position. This value has been measured via X-ray diffraction[2] and estimated via molecular dynamics simulations[3] to be in the range $E_{\mathrm{d}_0}$ =9-13 eV.

In irradiated p-type GaAs solar cells following a standardized annealing regime at 333 K[4], the remaining factors (RF) for the short-circuit current density $J_{\mathrm{SC}}$ can be fitted with the classical equation $\mathrm{RF}(J_{\mathrm{SC}}) = 1\text{-}C_x\log(1\text{+DDD}/D_x)$ for all electron energies in the range of interest if an adapted threshold energy of $E_{\mathrm{d}} = 21$ eV is used in the NIEL calculation, as shown in Fig. 1 (A). By using the same $E_{\mathrm{d}}$ value for the proton NIEL, the proton data collapse on the same degradation curve, which is then a pure function of the DDD only. The physical reason for the discrepancy between $E_{\mathrm{d}}$ and $E_{\mathrm{d}_0}$ was attributed[5] to the annihilation of defects formed by recoils with energy lower than 21 eV. The degradation behavior of the open-circuit voltage $V_{\mathrm{OC}}$ and the recombination current density $J_{02}$, however, differ significantly from the $J_{\mathrm{SC}}$ trend, showing that not all the solar cell parameters degrade the same way. In particular, two features are observed in GaAs cells: I) the proton DDD is more effective than the electron DDD in damaging $V_{\mathrm{OC}}$ and $J_{02}$, as visible in Fig. 1 (B) and (C); II) the adapted value $E_{\mathrm{d}} = 21$ eV leads to a good fit of the $V_{\mathrm{OC}}$ electron data only in the low-medium dose range, and to a poor fit of the $J_{02}$ electron data for all fluence values. For the latter parameter, a higher $E_{\mathrm{d}} = 35$ eV would result in a better fit of the high-energy electron spectrum above 1 MeV, as shown in Fig. 1 (D). With this $E_{\mathrm{d}}$ value, however, the NIEL of 0.5 MeV electrons would be virtually zero, in contradiction with the slight $J_{02}$ degradation observed for this electron energy, as visible in Fig. 1 (C).

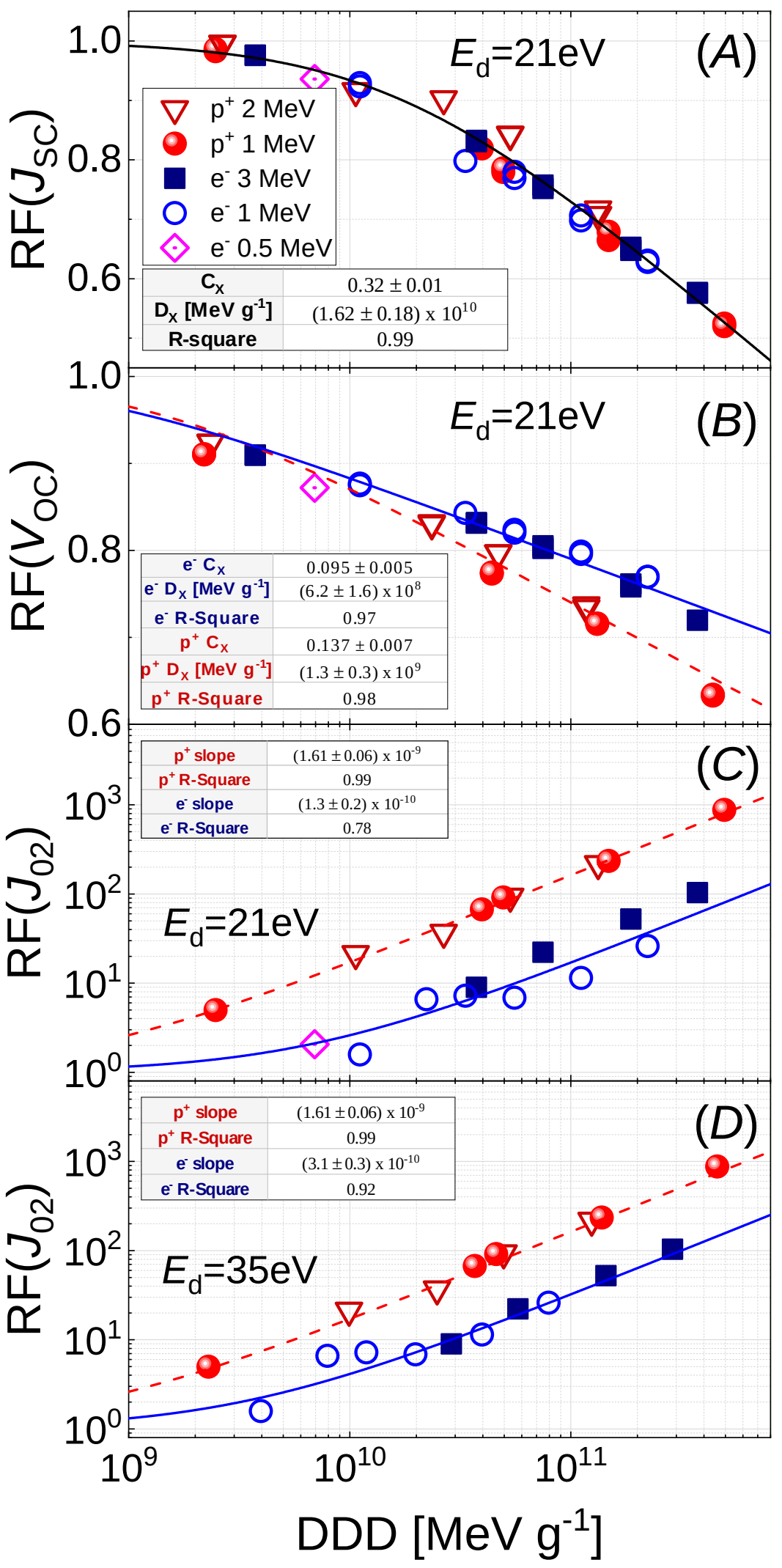

FIG. 1: GaAs component cell remaining factors of (A) short-circuit current density $J_{SC}$, (B) open-circuit voltage $V_{OC}$ and (C) recombination current density ($J_{02}$), with the DDD computed by using a threshold energy of $E_d = 21$ eV. (D) $J_{02}$ remaining factor with an adapted $E_d = 35$ eV. Further details on the cells can be found in section II.

Nevertheless, the adapted value $E_d = 21$ eV is generally used in the radiation analysis of GaAs-based solar cells. In the $V_{OC}$ and $J_{02}$ cases, empirical correction factors are defined as the ratio of the electron DDD ($D_e$) over the proton DDD ($D_p$) for a similar degradation level of the parameter, i.e. $R_{ep} = D_e/D_p$. The $R_{ep}$ factors allow $D_e$ to be converted in $D_p$ and vice versa. In p-type GaAs cells, it has been shown[1,6] that $R_{ep}^{V_{OC}} \approx 3$ and $R_{ep}^{J_{02}} \approx 9.5$.

A detailed knowledge of the defects produced by the displacement damage and their electrical effectiveness on the macroscopic cell parameters would allow the physical meaning behind the $R_{ep}$ factors and the adapted $E_d$ values adopted in the treatment of the solar cell remaining factors to be understood. Defect characterization in representative space component cells has been performed with deep-level transient spectroscopy (DLTS) on ad-hoc mesa diode structures[7]. A similar characterization of different solar cell technologies has been performed by several authors[8,9] by means of non-transient capacitance methods. In this work, admittance spectroscopy proved to be a suitable tool for defect characterization in standardized $2\,cm \times 2\,cm$ GaAs component cells irradiated with protons and electrons at different energies and fluences. The nature and electrical characteristics of four defects is probed in the p-type GaAs base, allowing their damaging effectiveness in terms of solar cell $J_{SC}$, $V_{OC}$ and $J_{02}$ to be estimated. The combined admittance-NIEL analysis suggests physical explanations for the values of $E_d$ and the $R_{ep}$ adopted in the DDD treatment of GaAs-based solar cell degradation data.

The paper is organized as follows: in Sec. II, the technology of the samples under test is presented, as well as details of the relevant irradiation experiments and measurement setups; in Sec. III, IV and V, the fitting routine of the solar cell admittance and the defect introduction rates is outlined in detail; in Sec. VI, the electrical characteristics of the irradiation-induced defect are presented, and conclusions are drawn in Sec. VII.

## II. EXPERIMENTAL DETAILS

The samples analyzed were $2\,cm \times 2\,cm$ bare GaAs $n^+/p$ component cells, fully representative of the middle sub-cell in lattice matched, 30% efficient, triple junction $Ga_{0.5}In_{0.5}P$/GaAs/Ge solar cells. The base layer was doped with a nominal acceptor concentration $N_a = 4\times10^{16}\,cm^{-3}$, whereas the emitter layer had a donor concentration of $N_d = 10^{18}\,cm^{-3}$. The cells featured surface field layers at the back (BSF) and at the front (FSF) side which were realized with higher band gap material. All relevant material parameters for GaAs were taken from literature data[10].

The solar cells were homogeneously irradiated at 300 K with 0.5, 1 and 3 MeV electrons and 1 and 2 MeV protons at different fluences $\Phi$. The irradiation history of all samples is summarized in Table I. Following irradiation, the solar cells were annealed in three steps: I) 48h under AM0 condition at $T = 300\,K$; II) 24h in dark conditions at $T = 333\,K$ and III) 24h in dark conditions at $T = 370\,K$. Annealing steps I and II were performed in accordance with the relevant ECSS space standards[4], whereas step III was performed to ensure stable conditions during the admittance mea-surements which were carried out up to this temperature.

The admittance of the solar cells was measured with an E4980a precision LCR meter in dark conditions, in the temperature range 190 K $< T <$ 370 K with steps of 10 K. The admittance was measured with a 0 V bias and an ac signal $\delta v$ with a root-mean-square (rms) value ranging from 10 to 30 $mV_{rms}$, in the frequency range 20 Hz $< f <$ 500 kHz. According to the Kramers-Kronig relations, only the capacitance data were used in the analysis. The accuracy of the capacitance measurements was better than 20 pF. The solar cell $J_{SC}$, $V_{OC}$ and $J_{02}$ data for these cells have already been measured and published in previous work[6].

TABLE I: Irradation history of GaAs Component Cells at $T$=300 K

| Particle | Energy [MeV] | Fluence [$cm^{-2}$] | Facility | *n.* of samples |
|---|---|---|---|---|
| $e^-$ | 0.5 | $1.4 \times 10^{16}$ | TU Delft | 2 |
| $e^-$ | 1 | $1 \times 10^{15}$ | TU Delft | 1 |
| $e^-$ | 1 | $2 \times 10^{15}$ | Ecole Polytechnique | 1 |
| $e^-$ | 1 | $3 \times 10^{15}$ | Ecole Polytechnique | 1 |
| $e^-$ | 1 | $5 \times 10^{15}$ | TU Delft | 2 |
| $e^-$ | 1 | $1 \times 10^{16}$ | TU Delft | 2 |
| $e^-$ | 1 | $2 \times 10^{16}$ | TU Delft | 2 |
| $e^-$ | 3 | $2 \times 10^{15}$ | TU Delft | 2 |
| $e^-$ | 3 | $5 \times 10^{15}$ | TU Delft | 2 |
| $e^-$ | 3 | $1 \times 10^{16}$ | TU Delft | 2 |
| $p^+$ | 1 | $8 \times 10^{11}$ | CSNSM Orsay | 1 |
| $p^+$ | 1 | $1 \times 10^{12}$ | CSNSM Orsay | 1 |
| $p^+$ | 1 | $3 \times 10^{12}$ | CSNSM Orsay | 2 |
| $p^+$ | 1 | $1 \times 10^{13}$ | CSNSM Orsay | 2 |
| $p^+$ | 2 | $4 \times 10^{11}$ | CSNSM Orsay | 1 |
| $p^+$ | 2 | $2 \times 10^{12}$ | CSNSM Orsay | 1 |
| $p^+$ | 2 | $5 \times 10^{12}$ | CSNSM Orsay | 2 |

## III. ADMITTANCE MODEL

The contribution of a discrete hole trap with homogeneous concentration $N_t$ to the space-charge region (SCR) capacitance $C_{SCR}$ under the stimulus of an external alternate bias $\delta v = v_{ac}\exp(2\pi f t)$ with amplitude $v_{ac}$ and frequency $f$ is modelled using the analytical approach proposed by Blood and Orton[11]. A deep level located at $E_t$ above the valence band $E_V$ is exposed to a hole flux $p \times v_{th}$, which is dependent on the position of the Fermi level $E_F$. Assuming an unitary degeneracy factor for the trap and negligible barrier lowering due to the Poole-Frenkel effect, the hole capture rate $c_p$ and the emission rate $e_p$ per unoccupied state are governed by the following relationships in thermal equilibrium:

$$c_p = \sigma_p v_{th} p = \sigma_p v_{th} N_V \exp\left(-\frac{E_F - E_V}{kT}\right);$$
$$e_p = c_p \exp\left(\frac{E_F - E_t}{kT}\right) = \sigma_p v_{th} N_V \exp\left(-\frac{E_t - E_V}{kT}\right). \tag{1}$$

$v_{th}$ is the rms velocity of holes in the valence band, $N_V$ is the density of states in the valence band and $\sigma_p$ denotes the hole capture cross-section. In the first half of the $\delta v$ positive cycle, the modulation $\delta E_F$ in the SCR leads to a trap-filling process with rate $c_p$ and a modulation of the defect charge state per unit area $\delta q_t$. In the second half of the positive cycle, the filled traps are emptied by hole emissions with rate $e_p$. The same applies to the negative ac cycle. If $\delta q_a$ denotes the charge variation due to the ionized acceptors $N_a$ at the edge of the SCR, the current oscillation per unit area in response to the applied ac bias is $\delta i = \mathrm{d}(\delta q_a + \delta q_t)/\mathrm{d}t$.

The charge modulation $\delta q_t$ responds directly to the driving bias $\delta v$, provided that $f \ll e_p/\pi$. In this low-frequency range, the trap is activated and the charge contribution to the overall capacitance is due to both $N_a$ and $N_t$: $C = C_{SCR} + \Delta C$. By making use of Poisson's equation in the SCR, the capacitance contribution $\Delta C$ associated with the deep level is evaluated as:

$$\frac{\Delta C}{C_{SCR}} = \frac{N_t}{N_a}\left(\frac{1 - \frac{x_t}{W_{SCR}}}{1 + \frac{N_t x_t}{N_a W_{SCR}}}\right), \tag{2}$$

where $W_{SCR}$ is the SCR width and $x_t$ is the position in the SCR where $E_t = E_F$. For $N_t \ll N_a$ Eq. (2) can be further simplified to:

$$\frac{\Delta C}{C_{SCR}} = \frac{N_t}{N_a}\sqrt{\frac{E_t - E_{F_p}}{qV_{bi}}}, \tag{3}$$

where $V_{bi}$ is the built-in voltage and $E_{F_p}$ the Fermi level in the p-side neutral region. At higher frequencies, the trap cannot emit holes during the emitting cycle since $f \gg e_p/\pi$, therefore $\delta q_t = 0$ and $C = C_{\text{SCR}}$ due solely to $N_{\text{a}}$.

A frequency-dependent expression for the total capacitance per unit area is obtained[11] from the complex part of $(2\pi f)^{-1}\,\delta i/\delta v$:

$$C(f) = C_{\text{SCR}} + \frac{\Delta C}{1+(f/f_{\text{t}})^2}. \tag{4}$$

Its derivation considers the change in majority carrier concentration in the SCR based on the combined response of the shallow acceptors as well as the trap to the applied bias. The occupation probability of the trap is calculated with the help of the capture and emission rates according to Eq. (1). The characteristic frequency $f_{\text{t}}$ defines the transition between the two states of the trap behavior and is derived as:

$$f_{\text{t}} = \frac{e_{\text{p}}}{\pi}\left(1+\frac{x_{\text{t}}N_{\text{t}}}{W_{\text{SCR}}N_{\text{a}}}\right). \tag{5}$$

The term in the brackets of Eq. (5) expresses the coupling between the response of the deep level and the shallow acceptors. For $N_{\text{t}} \ll N_{\text{a}}$ it can be neglected and the transition frequency is given by Eq. (1). If the temperature dependence of $f_{\text{t}}$ is known, an Arrhenius fit allows the trap energy level $E_{\text{t}}$ to be extracted, as well as the subsequent capture cross section $\sigma_{\text{p}}$, considering that the product $N_{\text{V}}\,v_{\text{th}}$ has a $T^2$ temperature dependence, whereas $\sigma_{\text{p}}$ is assumed to be temperature independent.

The $C$-$f$ characteristic exhibits a step-like behavior centered at $f = f_{\text{t}}(T)$ according to Eq. (4). At any given temperature, $f_{\text{t}}$ is the frequency point with maximum derivative and is easily identified in a measured spectrum as the peak of the functional $-f\text{d}C/\text{d}f$. The depletion capacitance $C_{\text{SCR}}$ is the value measured when ideally no deep level contributes and no carrier freeze-out occurs, i.e. $f \gg \max[e_{\text{p}}]$ and $f \ll \tau_{\varepsilon}^{-1}$, with $\tau_{\varepsilon}$ denoting the dielectric relaxation time of the dopant atoms in GaAs.

In the real case of $n$ deep levels characterized by a continuous distribution of available states in the band gap centered around the energy value $E_{\text{t}_0}$, the term $N_{\text{t}}$ in Eq. (3) is replaced by the total density of states $N_{\text{T}}(E)$. For each defect level, a Gaussian-shaped distribution is considered, characterized by a density function:

$$N_{\text{T,n}}(E) = \frac{N_{\text{t}_0\text{,n}}}{\lambda_{\text{n}}\sqrt{2\pi}}\exp\left[-\frac{1}{2}\left(\frac{E-E_{\text{t}_0\text{,n}}}{\lambda_{\text{n}}}\right)^2\right], \tag{6}$$

where $N_{\text{t}_0}$ is the total integrated concentration and $\lambda$ is the distribution width defining the localization of the electrons in the level. By replacing $N_{\text{t}}$ with $N_{\text{T}}(E)$ and $E_{\text{t}}$ with the variable $E$ in the previous equations, the admittance formalism presented so far is extended to the case of Gaussian distributions of states in the band gap. The total capacitance is thus obtained by substituting

Eq. (3) and (5) in Eq. (4) and integrating over the energy $E$ in the band gap. If all energy levels are not used in absolute terms, but are referenced to the valence band according to $E' = E - E_V$, the following expression is derived:

$$C = C_{SCR} + \sum_{k=1}^{n} \frac{C_{SCR}}{\lambda_k \sqrt{2\pi q V_{bi}}} \frac{N_{t0,k}}{N_a} \times$$

$$\times \int_{E'_{Fp}}^{E'_{Fp}+qV_{bi}} \frac{\sqrt{E' - E'_{Fp}} \exp\left[-\frac{1}{2}\left(\frac{E'-E'_{t0,k}}{\lambda_k}\right)^2\right]}{1+\left\{\frac{\pi f \exp\left[E'/(kT)\right]}{\sigma_{p,k} v_{th} N_V}\right\}^2} dE'. \quad (7)$$

The value $E'_{Fp}$ is obtained as $E'_{Fp} = E'_i - kT \ln(N_a/n_i)$, where $E'_i$ and $n_i$ are the intrinsic Fermi level position and the intrinsic carrier concentration, respectively. Deep levels outside the integral boundaries in Eq. (7) do not cross $E_F$ in the p-side SCR, and thus cannot contribute to the capacitance.

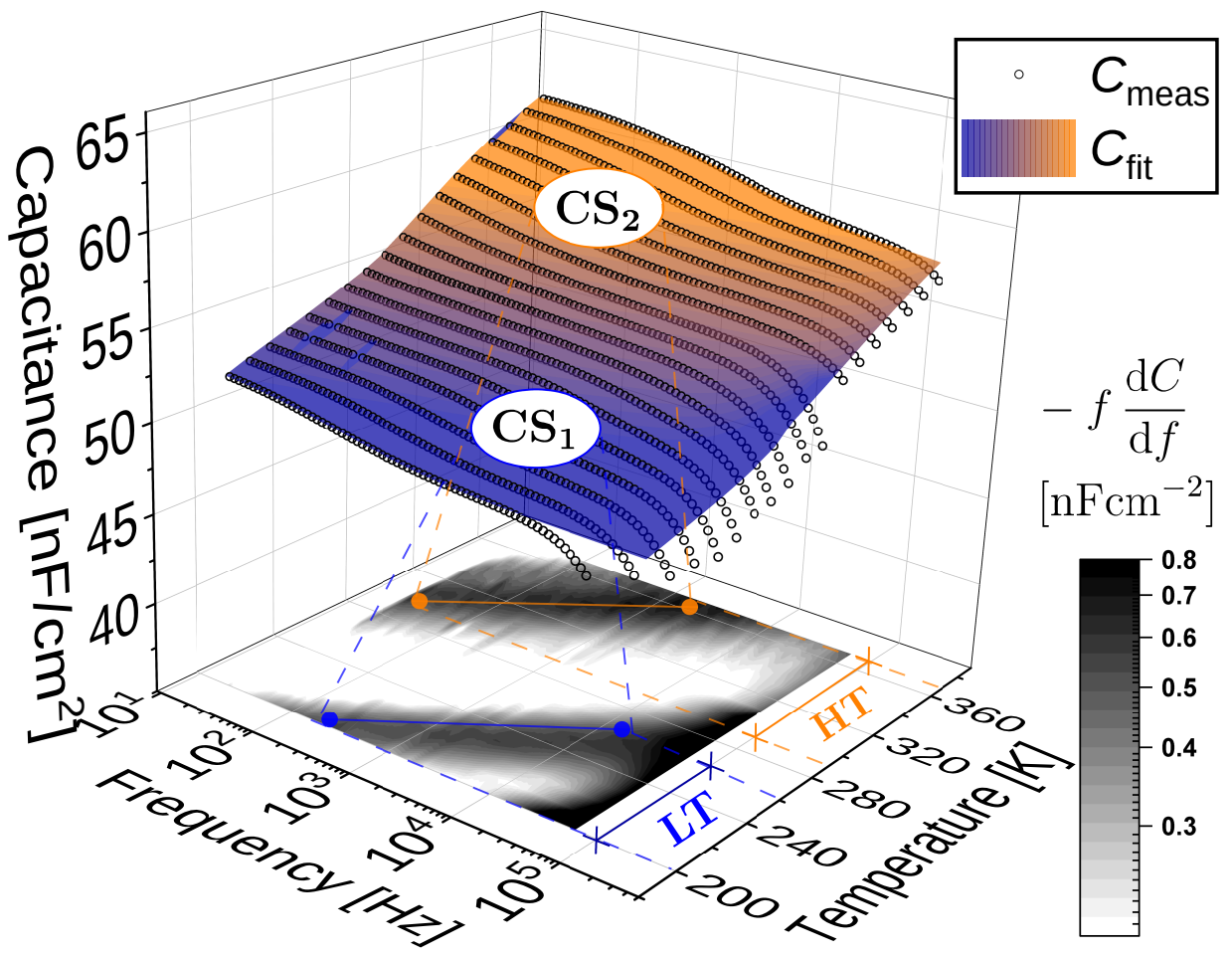

FIG. 2: $C$-$f$-$T$ spectra of p-type GaAs component cell irradiated with 3 MeV electron at a fluence of $5 \times 10^{15}$ cm$^{-2}$. Measured data points are represented by open circles, whereas the surface fit is obtained using Eq. (7). The surface color map distinguishes two distinct capacitance steps $CS_1$ (blue area) and $CS_2$ (orange area). On the $f$-$T$ plane, the projection of the function -$f$d$C$/d$f$ is illustrated.

## IV. FITTING PROCESS

### A. General considerations

A representative $C$-$f$-$T$ spectra for irradiated GaAs solar cell is illustrated in Fig. 2. Two main capacitance steps $CS_1$ and $CS_2$ are visible in the blue and orange area of the plot, respectively. On the $f$-$T$ plane, the projection of the function $-f\mathrm{d}C/\mathrm{d}f$ facilitates to pinpoint the $f_\mathrm{t}$-$T$ tracks. The evolution of $CS_1$ and $CS_2$ with temperature occurs separately in different areas of the $f_\mathrm{t}$-$T$ plot without significant overlap: $CS_1$ in the range $\mathrm{LT} = \{190\,\mathrm{K} < T < 260\,\mathrm{K}\}$ and $CS_2$ in the range $\mathrm{HT} = \{290\,\mathrm{K} < T < 370\,\mathrm{K}\}$. The drop-off at high frequencies is caused by the resonance of the total cell capacitance with the stray inductance of the external circuit. At low temperatures, the hole conduction at the BSF hetero-interface[12] results in a big capacitance drop. It limits the accessible frequency range and masks the effect of potential shallower levels below 190 K.

The fitting of the $C$-$f$-$T$ surface with Eq.(7) requires the knowledge of the exact value of $C_\mathrm{SCR}(T)$ from which in turn $N_\mathrm{a}(T)$ and later $N_{\mathrm{t}_0}$ can be accurately determined. Theoretically, $C_\mathrm{SCR}$ could be easily extracted for every temperature as the capacitance value measured in the high frequency range, although this – as mentioned before – has limited accessibility. Therefore, $C_\mathrm{SCR}$ should be included as a fitting parameter in Eq.(7). In addition, $C_\mathrm{SCR}$ is temperature-dependent due to the varying degree of ionization of the dopant atoms on both n- and p-type side of the junction. The temperature dependence is not a linear function over the wide temperature range investigated[13], thus leading potentially to an ambiguous fit. However, its value can be linearized separately in the temperature range LT and HT by using the expression $C_\mathrm{SCR}(T) = \alpha T + \beta$, with constants $\alpha$ and $\beta$ for each temperature regime.

The fitting process is thus split in two separate routines for the LT and the HT regime, leading to the determination of two separate values of $C_\mathrm{SCR}^\mathrm{LT}(T)$ and $C_\mathrm{SCR}^\mathrm{HT}(T)$ and avoiding ambiguities arising from single-temperature capacitance data. The splitting of the fitting process allows further simplification in the single routines. The defect responsible for the $CS_1$ step in the LT regime is always activated in the HT regime, i.e., the $f_\mathrm{t}$ associated to $CS_1$ shifts out of the measurable frequency range in the HT regime according to Eqs. (1) and (5). For this reason, the concentration of the defect responsible for $CS_1$ can be lumped together with $N_\mathrm{a}$ in the value of $C_\mathrm{SCR}^\mathrm{HT}$. This approach is in line with the treatment of all the defects independently from one another, and requires less computational resources. By embracing this approach, it should be noted that the resulting $C_{\mathrm{SCR}}^{\mathrm{HT}}$ includes the contribution of the $\mathrm{CS}_1$ defect concentration and $N_\mathrm{a}$, whereas $C_{\mathrm{SCR}}^{\mathrm{LT}}$ is only due to $N_\mathrm{a}$. In summary, the splitting of the fitting routine in the LT and HT range allows the temperature dependency of $C_{\mathrm{SCR}}$ to be linearized in the two regimes, thus leading to an unambiguous fit. Moreover, it improves the computational efficiency by reducing the number of defects required in Eq. (7) per each fitting routine.

## B. Parameter fitting

The number of defects $n$ included in each fitting routine is determined by the number of distinguishable peaks $f_\mathrm{t}$ found in the functional -$f\mathrm{d}C/\mathrm{d}f$: one single defect – labeled H1 – in the LT range, and three defects – labeled H2, H3 and H4 – in the HT range. The three defects in the HT range are not distinguishable in Fig. 2 because of the limitation in the resolution of the plot. The parameters to be determined are $E'_{\mathrm{t}_0}$, $N_{\mathrm{t}_0}$, $\sigma_\mathrm{p}$ and $\lambda$ for each defect, as well as the coefficients $\alpha$ and $\beta$ which determine $C_{\mathrm{SCR}}^{\mathrm{LT,RT}}(T)$ for the two temperature regimes. The fitting of the $C$-$f$-$T$ surface is performed in three steps:

(i) The trap signature $f_\mathrm{t}(T)$ is fitted with the Arrhenius equation Eqs. (5),(1), yielding $E'_{\mathrm{t}_0}$ and an initial value of $\sigma_\mathrm{p}$.

(ii) The initial values for $\alpha$, $\beta$, $N_{\mathrm{t}_0}$ and $\lambda$ are chosen manually according to their respective influences on the $C$-$f$ curves.

(iii) The refined values for $\sigma_\mathrm{p}$, $N_{\mathrm{t}_0}$ and $\lambda$, as well as the coefficients $\alpha$ and $\beta$ are obtained with the help of the algorithm that minimize the residual function $S_{\mathrm{rms}}$:

$$S_{\mathrm{rms}} = \sqrt{N_{\mathrm{point}}^{-1} \sum_T \sum_f (C_{\mathrm{meas}} - C_{\mathrm{fit}})^2}, \tag{8}$$

where $N_{\mathrm{point}}$ is the number of useful data points.

The results of the fitting process for all cells, particle energies and fluences is summarized in Table II. For $E'_{\mathrm{t}_0}$, the uncertainty range of the Arrhenius fit is indicated. The uncertainty associated with the remaining fitting parameters is obtained with a Monte Carlo approach[14], and represents the 95% confidence interval obtained over a set of 100 routines, where the starting values for each parameter is varied randomly in a range of two order of magnitude with respect to the expected value. It is useful for the manual determination of the starting parameters in step (ii) to consider the contribution of each fitting parameter in the step-like shape of the $C$-$f$ curve: $C_{\mathrm{SCR}}^{\mathrm{LT,HT}}$ is the baseline capacitance signal, which is easily accessible in the high-frequency range on the lower temperature curves for each regime. The linearization of $C_{\mathrm{SCR}}^{\mathrm{LT,HT}}(T)$ ensures coherent values also at the temperatures where the high-frequency capacitance is not accessible due to the aforementioned limitations. The steepness of the capacitance step is solely determined by the parameter $\lambda$ in Eq. (7). The amplitude of the capacitance step is thus mainly determined by the parameter $N_{\mathrm{t0}}$. The $f_{\mathrm{t}}$ position is determined by $E_{\mathrm{t0}}$ and $\sigma_{\mathrm{p}}$ according to Eq. (5) with $N_{\mathrm{t0}} \ll N_{\mathrm{a}}$. If $E_{\mathrm{t0}}$ is fixed by the Arrhenius fit, the position of $f_{\mathrm{t}}$ is solely determined by $\sigma_{\mathrm{p}}$. It should be mentioned that, as $E_{\mathrm{t0}}$ is fixed for all the irradiated cells, a slight variation within the uncertainty of the energy level is compensated by a large variation of $\sigma_{\mathrm{p}}$ according to Eq. (1), thus explaining the high uncertainty of this parameter.

Taking into account the accuracy of the LCR meter and the uncertainty of the doping levels of the cells, a value of $N_{\mathrm{t0}} \approx 8\times10^{13}\,\mathrm{cm}^{-2}$ represents the minimum concentration sensitivity, resulting in a detection factor $N_{\mathrm{t0}}/N_{\mathrm{a}} \approx 2\times10^{-3}$. It has to be noted that $CS_1$ and $CS_2$, ascribed to the effect of irradiation-induced defects, have been checked against other possible causes, e.g., resistance effects, dielectric relaxation and hetero-interface barriers.

The effect of all defects on the cell capacitance is distributed over the entire $f$-$T$ domain. It is possible to produce a single plot, which condenses all data of the two $f$-$T$ domains in a unique curve, so that the effect of all the defects is visible at a glance. This can be achieved by plotting -$f\mathrm{d}C/\mathrm{d}f$ over a scaled frequency axis computed as $fT^2 f_{\mathrm{o}} \exp(E_{\mathrm{f}}/kT)$, where $f_{\mathrm{o}}$ is a constant and the value $E_{\mathrm{f}}$ is an exponential scaling factor, which equals the activation energy of H2 for the spectra in the HT temperature range and of H1 in the LT temperature range. The resulting plot is illustrated in Fig. 3 (A), and the single contributions of H2, H3 and H4 in the HT range can be differentiated in the low scaled-frequency range. In addition, the density of states in the GaAs band gap extracted from the fitting process is depicted in Fig. 3 (B).

## V. NIEL DEFECT ANALYSIS

### A. NIEL Calculation

The defect concentration data obtained from the admittance fitting routine are correlated with the displacements created by the primary particle of energy $\xi$ in the crystal through all recoils.

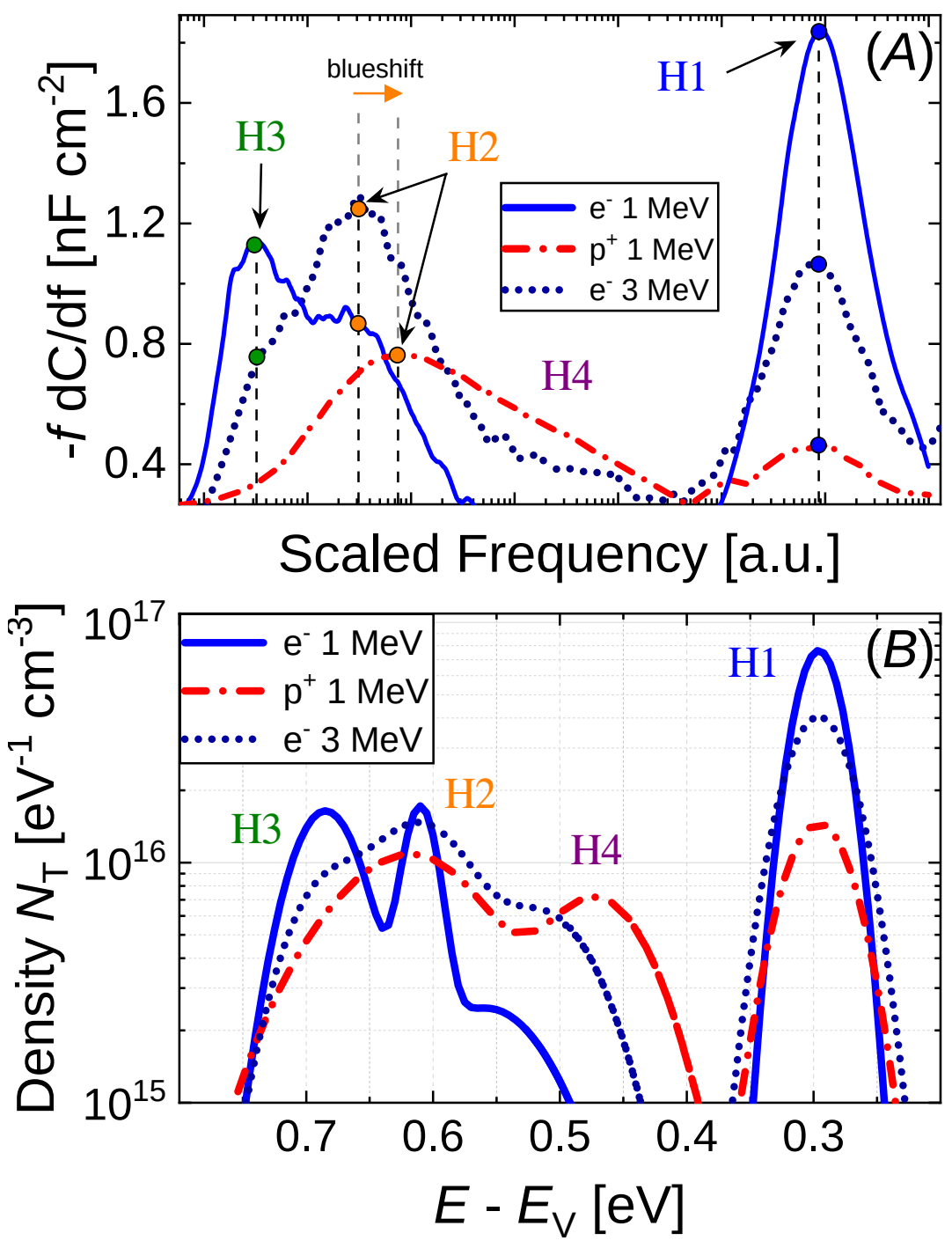

FIG. 3: (A) Scaled admittance signal $-f\mathrm{d}C/\mathrm{d}f$, and (B) combined density of states in the band gap of p-type GaAs component cells irradiated with a similar amount of DDD introduced by 1 and 3 MeV electrons and 1 MeV protons.

Under the assumption that the particle slow-down within the layers is negligible, the DDD is calculated as:

$$\mathrm{DDD} = \int_{\xi} \mathrm{NIEL}(\xi)\phi(\xi)\mathrm{d}\xi \,. \tag{9}$$

The particle NIEL in the GaAs compound is estimated by assuming a linear combination of the NIEL of the individual elements, weighted according to their atomic weight (Bragg's rule). For the single element, the NIEL is computed analytically as:

$$\mathrm{NIEL}(\xi) = \frac{N}{A}\int_{E_\mathrm{d}}^{E_\mathrm{max}} E_\mathrm{R} L(E_\mathrm{R}) \frac{\partial \sigma_\mathrm{Cs}(\xi, E_\mathrm{R})}{\partial E_\mathrm{R}} \mathrm{d}E_\mathrm{R}, \tag{10}$$

where $N$ is the Avogadro number and $A$ is the atomic mass. The integration variable $E_\mathrm{R}$ is the recoil energy. $E_\mathrm{d}$ is minimum energy transferred to a recoil that results in Ga and As displacements. In the following sections, the NIEL and related DDD are marked with the subscript "0" when the value $E_\mathrm{d} = E_{\mathrm{d}_0} = 10\,\mathrm{eV}$ is adopted. $E_\mathrm{max}$ is the maximum energy that the particle can transfer to the recoil. $\partial\sigma_\mathrm{Cs}/\partial E_\mathrm{R}$ is the differential cross-section for Coulomb scattering, quantifying the probability that the primary particle of energy $\xi$ generates a recoil with energy $E_\mathrm{R}$. $L(E_\mathrm{R})$ is the recoil partition function, which quantifies the fraction of $E_\mathrm{R}$ that goes into displacements. It is computed by means of the Robinson formulation[15] of the Lindhard partition function[16]. For the electron $\partial\sigma_{Cs}/\partial E_R$, the Mott differential cross section[17] with the approximation proposed by Boschini *et al.* is adopted[18], accounting for screened Coulomb fields, finite size and rest mass of the target nuclei. For protons, the Wentzel–Molière expression[19] is used. The resulting NIEL values are similar to the one calculated via the online SR-NIEL calculator[20]. The $NIEL_0$ contribution factor $\chi(E)$, defined as:

$$\chi(E) = \frac{\int_{E_{d_0}}^{E} E_R \, L \, \frac{\partial\sigma_{Cs}}{\partial E_R} \mathrm{d}E_R}{\int_{E_{d_0}}^{E_{max}} E_R \, L \, \frac{\partial\sigma_{Cs}}{\partial E_R} \mathrm{d}E_R}, \quad (11)$$

quantifies the $NIEL_0$ contribution of the recoil spectrum from $E_{d_0}$ up to the energy $E$. From the plots of the $\chi(E)$ factor, reported as a percentage in Fig. 4, it is possible to identify the effective recoil spectrum which deposits the DDD for each irradiating particle. For instance, 80% of the damage caused by 1 MeV electrons is due to recoils with energy below 40 eV. The same recoil spectrum accounts for only 40% of the damage in the 3 MeV electron case, and only for 10% of the damage in the 1 MeV proton case. The recoil spectrum triggered by the irradiating particle has several implications on the formation of the defects within the lattice. Recoils with energy just above $E_{\mathrm{d}_0}$ can only create one or no additional displacement. Therefore, the displacements produced by 1 MeV electrons, which can trigger recoils up to few tens of eV, are generated in a region of the lattice not populated by many other neighboring point defects. The displaced atoms are most likely isolated – i.e., they do not interact with other closely spaced lattice imperfection. In the proton case, on the other hand, recoils with much higher energy have a large contribution in the $\mathrm{NIEL}_0$. A high-energy recoil can trigger a long-range displacement cascade resulting in the formation of highly-damaged regions[21] where closely-spaced point defects interact and form stable clusters. The size of these clusters can extend up to several GaAs lattice constants[22].

TABLE II: Defect parameters resulting from the fitting routine of the GaAs component cells. The first column indicates the irradiation particle, energy (in MeV) and fluence (in $cm^{-2}$).

| cell | H1 | | | H2 | | | H3 | | | H4 | | |
|---|---|---|---|---|---|---|---|---|---|---|---|---|
| | $E'_{t_0} = (0.29 \pm 0.01)$ eV | | | $E'_{t_0} = (0.61 \pm 0.02)$ eV | | | $E'_{t_0} = (0.68 \pm 0.03)$ eV | | | 0.45 eV $< E'_{t_0} <$ 0.55 eV | | |
| | $N_{t_0}$ | $\sigma_p$ | $\lambda$ | $N_{t_0}$ | $\sigma_p$ | $\lambda$ | $N_{t_0}$ | $\sigma_p$ | $\lambda$ | $N_{t_0}$ | $\sigma_p$ | $\lambda$ |
| | [$cm^{-3}$] | [$cm^{-2}$] | [eV] | [$cm^{-3}$] | [$cm^{-2}$] | [eV] | [$cm^{-3}$] | [$cm^{-2}$] | [eV] | [$cm^{-3}$] | [$cm^{-2}$] | [eV] |
| | $\times 10^{14}$ | $\times 10^{-16}$ | $\times 10^{-3}$ | $\times 10^{14}$ | $\times 10^{-16}$ | $\times 10^{-3}$ | $\times 10^{14}$ | $\times 10^{-16}$ | $\times 10^{-3}$ | $\times 10^{14}$ | $\times 10^{-16}$ | $\times 10^{-3}$ |
| e.5 $1.4\times 10^{16}$ | $0.8 \pm 0.3$ | $2 \pm 1.4$ | $20 \pm 9$ | – | – | – | – | – | – | – | – | – |
| e1 $1\times 10^{15}$ | $1.5 \pm 0.6$ | $4.7 \pm 3.4$ | $33 \pm 15$ | – | – | – | – | – | – | – | – | – |
| e1 $3\times 10^{15}$ | $5.8 \pm 1.6$ | $3.4 \pm 2.5$ | $14 \pm 7.4$ | $0.6 \pm 0.4$ | $40 \pm 28$ | $19 \pm 19$ | $2.5 \pm 0.7$ | $20 \pm 30$ | $36 \pm 38$ | $0.5 \pm 0.3$ | 40 | $78 \pm 61$ |
| e1 $5\times 10^{15}$ | $9.1 \pm 1.6$ | $4.5 \pm 2.5$ | $25 \pm 7$ | $0.9 \pm 0.5$ | $109 \pm 80$ | $10 \pm 10$ | $3.9 \pm 0.6$ | $80 \pm 60$ | $46 \pm 24$ | $0.7 \pm 0.2$ | 109 | $33 \pm 17$ |
| e1 $1\times 10^{16}$ | $16 \pm 3$ | $3.7 \pm 1.6$ | $18 \pm 3$ | $1.7 \pm 0.3$ | $104 \pm 44$ | $11 \pm 3$ | $6.9 \pm 1.2$ | $83 \pm 35$ | $35 \pm 6$ | $1.8 \pm 0.4$ | 104 | $50 \pm 10$ |
| e1 $2\times 10^{16}$ | $33 \pm 6$ | $3.5 \pm 1.4$ | $17 \pm 3$ | $4.8 \pm 0.8$ | $82 \pm 45$ | $12 \pm 4$ | $11 \pm 2$ | $77 \pm 42$ | $26 \pm 5$ | $3 \pm 0.5$ | 82 | $49 \pm 8$ |
| e3 $2\times 10^{15}$ | $10 \pm 3$ | $4 \pm 1.7$ | $34 \pm 6$ | $4 \pm 0.5$ | $67 \pm 28$ | $29 \pm 5$ | $2.8 \pm 0.6$ | $45 \pm 19$ | $26 \pm 4$ | $2 \pm 0.2$ | 67 | $40 \pm 10$ |
| e3 $5\times 10^{15}$ | $25 \pm 3$ | $4.7 \pm 1.5$ | $32 \pm 5$ | $12 \pm 1.4$ | $72 \pm 30$ | $36 \pm 5$ | $5.7 \pm 0.7$ | $57 \pm 24$ | $31 \pm 5$ | $6 \pm 0.7$ | 72 | $40 \pm 3$ |
| e3 $1\times 10^{16}$ | $58 \pm 5$ | $3.4 \pm 0.9$ | $14 \pm 4$ | $28 \pm 3.3$ | $76 \pm 16$ | $35 \pm 3$ | $10 \pm 1.3$ | $57 \pm 12$ | $31 \pm 3$ | $10 \pm 1.2$ | 76 | $40 \pm 4$ |
| p1 $8\times 10^{11}$ | $3.6 \pm 1.3$ | $4.9 \pm 3.6$ | $45 \pm 8$ | – | – | – | – | – | – | – | – | – |
| p1 $1\times 10^{12}$ | $4.6 \pm 1.3$ | $4.9 \pm 3.6$ | $41 \pm 8$ | $3 \pm 0.8$ | $191 \pm 111$ | $44 \pm 6$ | $1.5 \pm 0.4$ | $54 \pm 39$ | $47 \pm 6$ | $2.5 \pm 0.7$ | 191 | $50 \pm 6$ |
| p1 $3\times 10^{12}$ | $9.5 \pm 2$ | $3.9 \pm 1.6$ | $26 \pm 8$ | $12 \pm 2.3$ | $221 \pm 93$ | $49 \pm 7$ | $3.3 \pm 0.6$ | $50 \pm 21$ | $46 \pm 6$ | $7 \pm 1.3$ | 221 | $40 \pm 5$ |
| p1 $1\times 10^{13}$ | $34 \pm 3.6$ | $3.2 \pm 1$ | $39 \pm 8$ | $45 \pm 2.2$ | $224 \pm 51$ | $65 \pm 8$ | $5.4 \pm 0.6$ | $25 \pm 21$ | $35 \pm 6$ | $16 \pm 0.7$ | 224 | $30 \pm 4$ |
| p2 $2\times 10^{12}$ | – | – | – | $3.5 \pm 0.7$ | $138 \pm 51$ | $70 \pm 10$ | $0.2 \pm 0.2$ | $125 \pm 55$ | $26 \pm 4$ | $1.9 \pm 0.3$ | 138 | $84 \pm 25$ |
| p1 $5\times 10^{12}$ | $10 \pm 6.9$ | $4 \pm 2.5$ | $61 \pm 8$ | $9.7 \pm 0.3$ | $147 \pm 68$ | $71 \pm 10$ | $0.6 \pm 0.5$ | $14 \pm 10$ | $36 \pm 20$ | $4 \pm 1.3$ | 147 | $40 \pm 25$ |

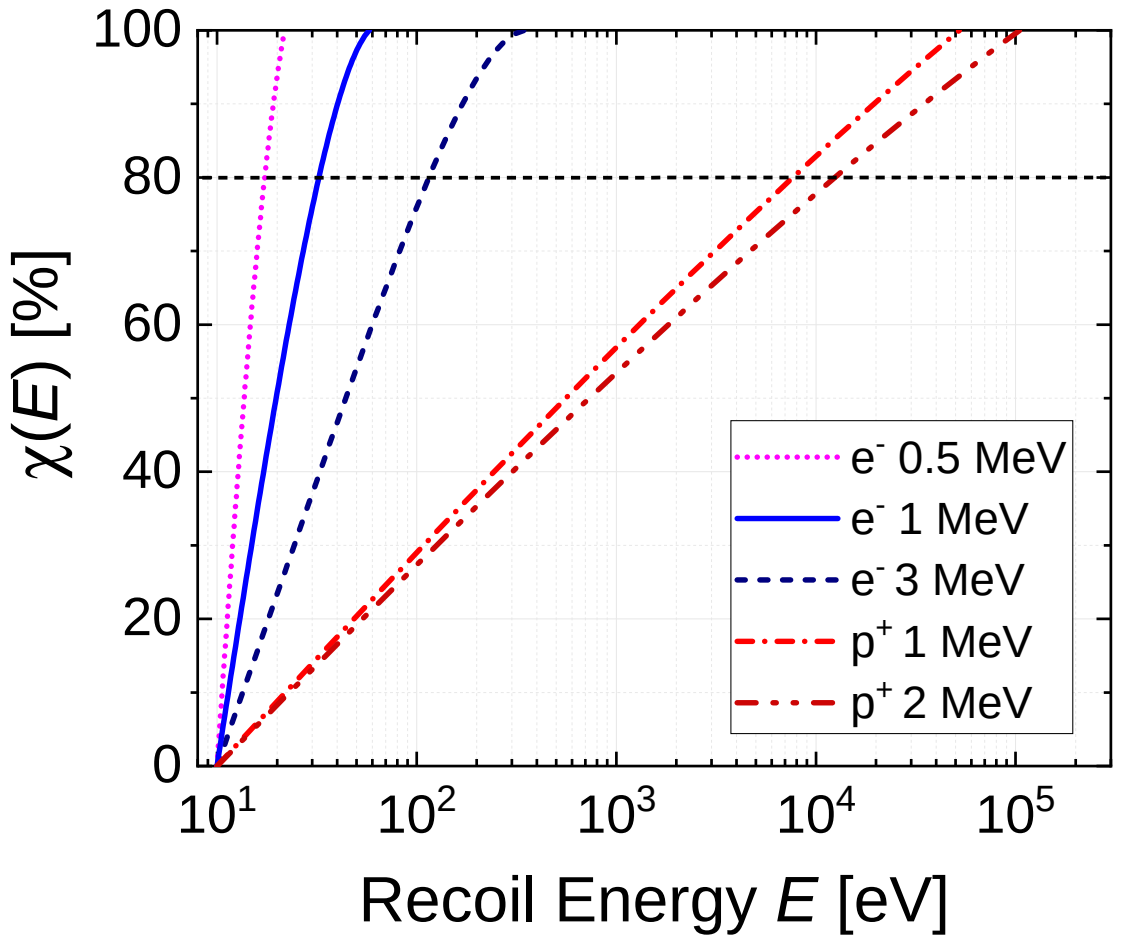

FIG. 4: Relative NIEL contribution $\chi(E)$ originating from recoils with energies below $E$ for 0.5, 1 and 3 MeV electrons and 1 and 2 MeV protons in GaAs.

### B. Fitting of the Defect $E_\mathrm{d}$

The admittance measurements yield individual defect concentrations as a function of different particle energies and fluences. Therefore, it is a straightforward approach to see whether they can be collapsed onto a single characteristic $N_{\mathrm{t}_0}$-DDD curve. It will be shown in the following section that the data fall onto a single straight line if an adapted $E_\mathrm{d}$ in the electron case is used. The value $E_\mathrm{d}$ can be directly related to the physical defect configuration. It represents the minimum recoil energy whose displacements contribute effectively to the final defect structure. When $E_\mathrm{d}$ approaches the threshold energy for Ga and As displacement $E_{d_0}$, the defect unit cell is made up of a single-displacement structure, i.e., a simple Frenkel pairs. When $E_d > E_{d_0}$, on the other hand, the defect unit cell is composed of a more complex structure, such as defects with higher formation energy (e.g., antisites) or multi-displacement defects (e.g., divacancies). The fitting of $E_d$ for a given defect such that the defect introduction rates fall on the same NIEL curve for all electron energies is therefore a way to draw conclusions on the physical defect configuration in the lattice directly from admittance measurements. Only the electron data are used for this procedure due to the higher sensitivity of the electron NIEL to $E_d$ compared to the proton NIEL in the energy range of interest. It has to be noted that the NIEL is based on a purely analytical calculation. The adapted threshold energy for the defect in question is a way to achieve the correct energy scaling with regard to the introduction rate of different particles. No conclusion on relative contributions of other defects in the displacement damage can be drawn from this fitting alone.

The electron displacement introduction rate (eDIR) is defined as $\text{eDIR} = N_{t_0}/D_e$ in analogy to the defect introduction rate $k_t = N_{t_0}/\Phi$. Likewise, pDIR is defined with respect to the adapted proton dose. When eDIR and pDIR are different, the different particle effectiveness is accounted for by the factor $\text{R}^{\text{d}}_{\text{ep}} = \text{pDIR/eDIR}$, defined in analogy to the electrical $\text{R}_{\text{ep}}$. The defect $\text{R}^{\text{d}}_{\text{ep}}$ allows $D_e$ to be scaled so that the $N_{t_0}$-$\text{D}_\text{e}$ plot falls onto the $N_{t_0}$-$\text{D}_\text{p}$ plot.

While a similar approach has also been followed in the past for the solar cell macroscopic electric parameters, any conclusions in this case were hampered by the fact that only the combined behavior of all defects, weighted by their electrical effectiveness, could be studied. Finally, it has to be noted that since the cells were irradiated at 300 K and all measurements were performed after 370 K annealing, an annealed defect configuration was probed in all cases.

## VI. IRRADIATION-INDUCED DEFECTS

### A. H1

The capacitance step $\text{CS}_1$ is characterized by a clear, distinguishable, single peak in the $-f\,\text{d}C/\text{d}f$ functional, thus allowing an easy determination of $f_t(T)$. By making use of Eq. (5), the energy level associated with the defect, labeled H1, is found: $E'_{t_0} = 0.29$ eV. The fitting of H1 introduction rates with the electron NIEL yields $E_d = 21$ eV, which suggests that the defect is either formed by isolated antisites[23] or a different double-displacement defect. A similar energy level in irradiated p-type GaAs has been already reported in literature[24,25]. The $N_{t_0}$-DDD and the introduction rate plots are reported in Fig. 5 (A) and (B). In Fig. 5 (B), the NIELs computed with different $E_d$ values are superimposed in order to illustrate the sensitivity of the $E_d$ fitting process.

eDIR and pDIR coefficients of $1.4\times10^4$ and $7.1\times10^3$ g cm$^{-3}$ MeV$^{-1}$ result, respectively, thus yielding $R^d_{ep} = 0.5$. The electron dose $D_e$ is twice as effective as the corresponding proton $D_p$ in producing H1.

### B. H2 and H3

The capacitance step $CS_2$ for electron-irradiated cells is characterized by the presence of two overlapping peaks in the $-f\mathrm{d}f/\mathrm{d}C$ functional, which are attributed to two separate irradiation-induced defects, H2 and H3. The contributions of H2 and H3 are de-coupled by performing a Gaussian de-convolution on the $-f\mathrm{d}f/\mathrm{d}C$ data. In this way, the single $f_t(T)$ are extracted and fitted with Eq. (5), leading to two energy levels $E'_{t_0} = 0.61$ eV and $E'_{t_0} = 0.68$ eV for H2 and H3, respectively. In the proton case, on the other hand, the peak associated with H3 is only detectable in the higher fluence cases. The NIEL analysis on the electron data yields $E_d$ values for H2 and H3 of 38 eV and 9 eV respectively. The low $E_d$ value for H3 suggests that it can be ascribed to a single-displacement defect, i.e., a Frenkel pair. On the other hand, H2 can be ascribed to a more complex defect structure. The energy level of H2 has been associated in the literature to the single donor configuration of the EL2 defect[26,27], possibly due to As antisites ($As_{Ga}$) or more complex structures involving $As_{Ga}$ coupled with As interstitial.

The $N_{t_0}$-DDD plots are reported in Fig. 5 (C), whereas the eDIR and $R^d_{ep}$ factor are reported in Fig. 5 (D). The defect H3 exhibits an $R^d_{ep} = 0.38$, whereas $R_{ep} = 1$ is found for H2. A begin-of-life H3 concentration of about $(9\pm2)\times10^{13}$ cm$^{-3}$ is obtained in the linear fit of both the $N_{t_0}$-$D_e$ and $N_{t_0}$-$D_p$ plots. This indicated that H3 may be present in the material already prior irradiation.

### C. H4

A small shoulder characterizes the H2 peak, as visible in Fig. 3 (A) for the proton and 3 MeV electron cases. This capacitance contribution is attributed to the deep level H4. Since only a featureless peak is detected in the $-f\mathrm{d}C/\mathrm{d}f$ plots, it is not possible to determine the trap $f_t(T)$. The H4 contribution is included in the admittance model by using a fixed value of $\sigma_p$ as the one of H2. In this way, the fitting routine yields the apparent H4 energy level, which shows variations in the range $0.45\,\text{eV} < E^*_{t_o} < 0.55\,\text{eV}$ for the different samples analyzed. The NIEL analysis yields a value of $E_d = 33\,\text{eV}$ and $\text{R}^{\text{d}}_{\text{ep}} = 1$.

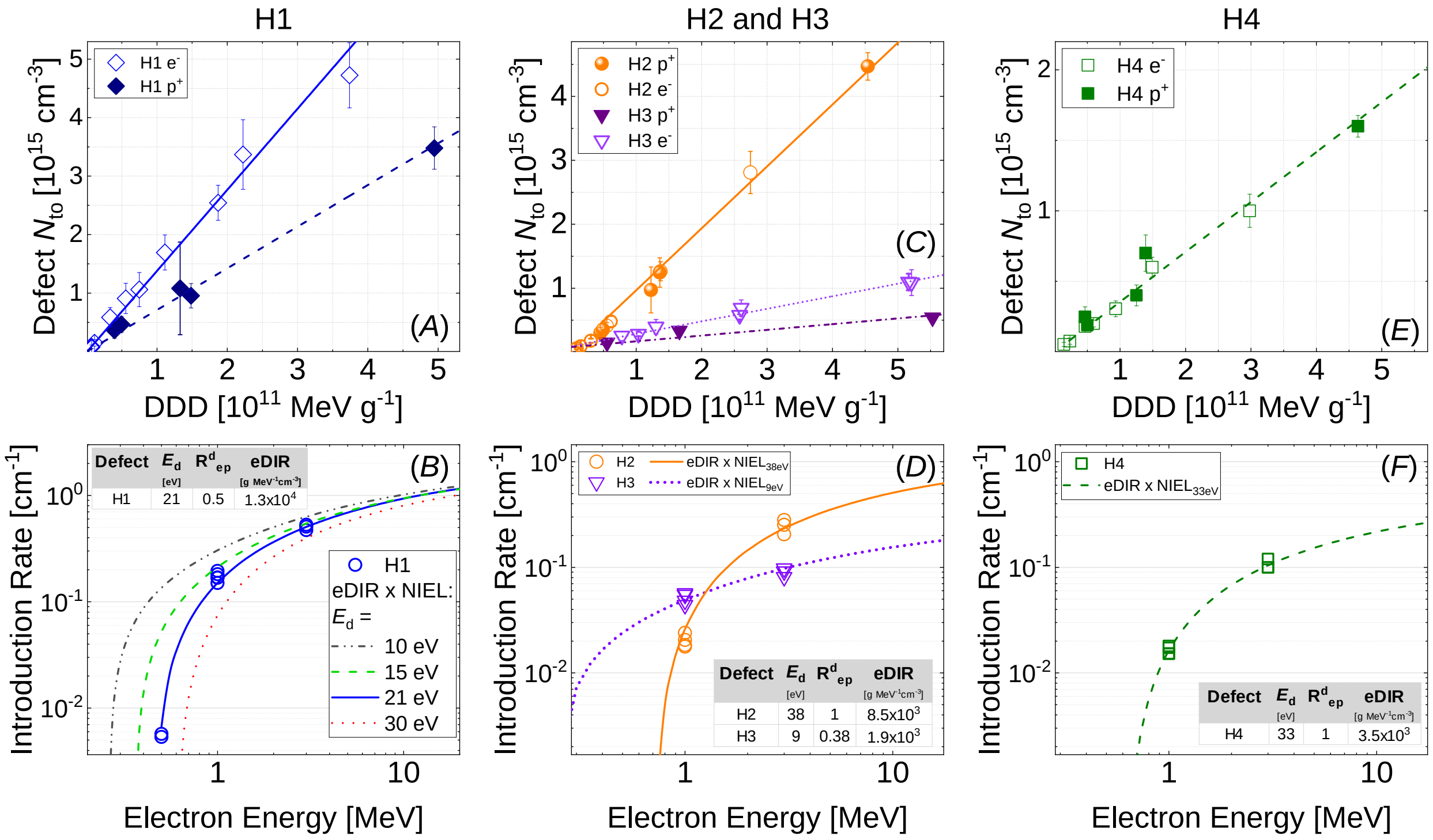

FIG. 5: Defect concentration as a function of adapted DDD and introduction rate as a function of particle energy for (A,B) H1, (C,D) H2 and H3, and (E,F) H4 in p-type GaAs component cells. The adapted DDD for each defect is computed by using the appropriate $E_d$ values, reported in the insets in (B), (D) and (F).

## VII. DISCUSSION

### A. Effective NIEL*

With the knowledge of the absolute number of defects generated and the associated defect energy $E_d$, the total displacement damage can be calculated for each irradiating particle type and energy. The actual NIEL introduced, denoted as NIEL*, can be computed by taking into account each single defect introduction rate with the associated threshold energy:

$$\mathrm{NIEL}^{*} = \rho^{-1} \sum_{i=1}^{n} (k_{\mathrm{t}} \times E_{\mathrm{d}})_i \,, \tag{12}$$

where $\rho$ is the material density and $n$ is the total number of type of defects introduced into the material; in this case, $n = 4$. The results for H1-H4 are reported in Fig. 6. The theoretical $\mathrm{NIEL}_0$ curves, computed using $E_{\mathrm{d}_0} = 10\,\mathrm{eV}$, are also superimposed using solid lines. The measured $\mathrm{NIEL}^{*}$ values deviate from the theoretical $\mathrm{NIEL}_0$ by a factor of 10. The missing $\mathrm{NIEL}^{*}$ contribution responsible for this gap is expected to arise from defects that cannot be gauged via admittance spectroscopy in p-type samples, i.e., electron traps, defects located below $E_{\mathrm{F_p}}$ or whose energy level does not lie in the forbidden band gap. A complementary analysis on at least the electron traps located in the upper half of the band gap is required to arrive to a complete estimation of the $\mathrm{NIEL}^{*}$. By considering only the introduction rates of the electron traps E1-E5 detected in irradiated n-type GaAs cells (1 MeV $e^-$ data from Pons *et al.*[28]), an approximated $\mathrm{NIEL}^{*}$ close to $\mathrm{NIEL}_0$ can be estimated. It has been demonstrated in the past that these traps are also introduced in p-type GaAs[27]. Therefore, assuming comparable introduction rates of E1-E5 for n- and p-type material, the bulk of the gap between $\mathrm{NIEL}^{*}$ and $\mathrm{NIEL}_0$ is due to the missing contribution of E1-E5 in Eq. (12).

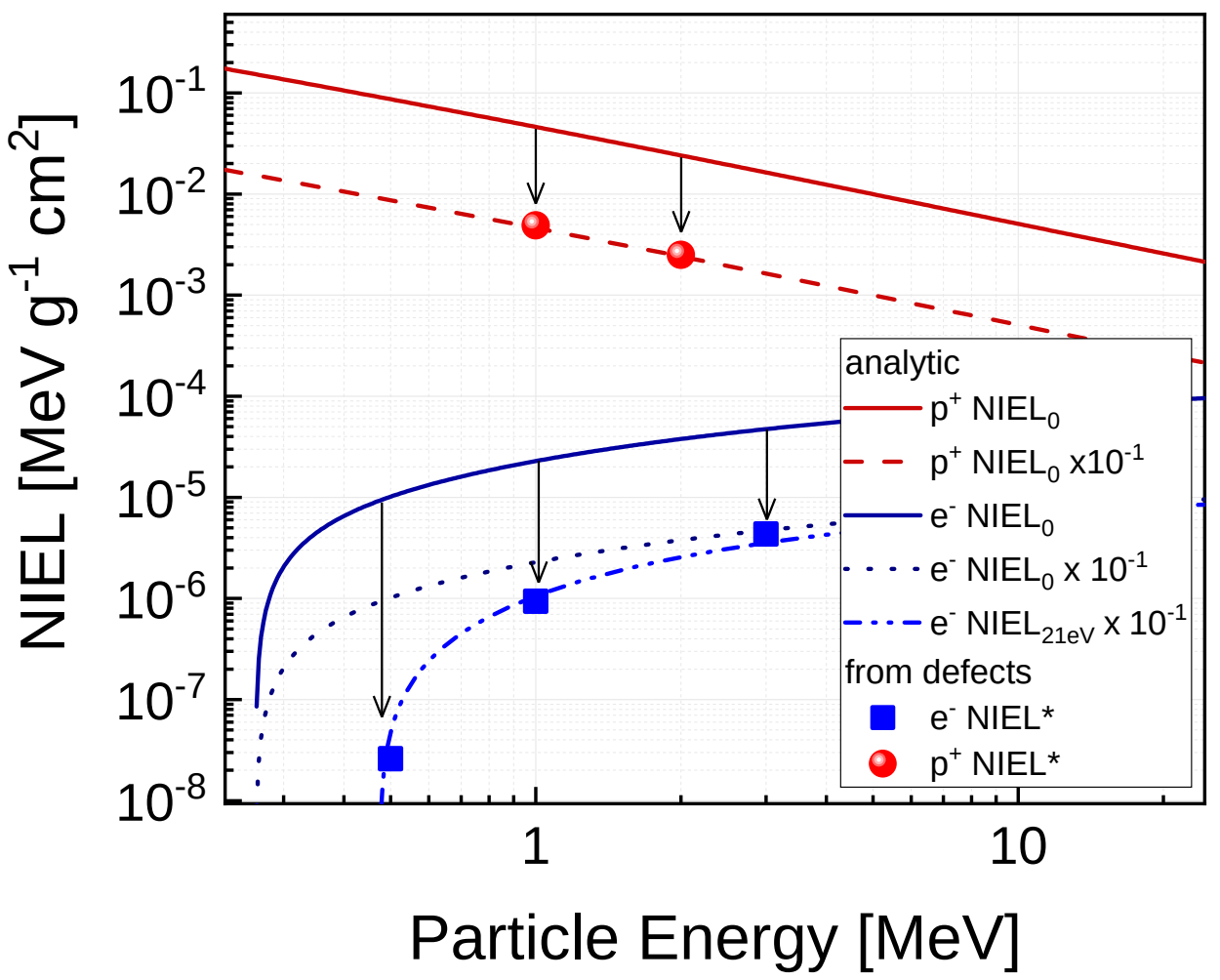

FIG. 6: Measured (data points) effective $\mathrm{NIEL}^{*}$ in proton and electron irradiated p-type GaAs computed with Eq. (11). The theoretical NIEL (solid curves) are calculated according to Eq. (10) with different threshold energies and superimposed to the plot.

Furthermore, it should be mentioned that the $\mathrm{NIEL}_0$ value accounts for the non-ionizing energy deposited by irradiating the material at 0 K. It has been already recognized that the theoretical

NIEL is not necessarily proportional to the total number of defects formed after irradiation[21] because of non-linear processes taking place in the lattice, e.g., the formation of clustered regions and the defect annihilation during both the irradiation experiment and the post annealing treatment[29]. Although a general conclusion cannot be drawn based solely on the data available from this analysis, it is remarkable how the NIEL$^*$ relative to H1-H4 falls on an attenuated NIEL $\times 10^{-1}$ curve computed with $E_\mathrm{d} = 21$ eV, as shown in Fig. 6. Despite the fact that the H1 defect with $E_\mathrm{d} = 21$ eV is the most prevalent defect, this $E_\mathrm{d}$ value is thus not representative of any particular defect structure, but rather an empirical value with which it is possible to describe the introduction of the effective NIEL$^*$ in p-type GaAs in the form of stable defects H1-H4. It follows that the reference NIEL calculated with $E_\mathrm{d} = 21$ eV can be used to compare a similar level of damage in terms of H1-H4 defects for cells irradiated with different particles and energies. Since no cell electrical parameters were involved in this derivation, it can be further concluded that this threshold energy is a GaAs material-specific parameter, independent of the detailed electrical device design.

### B. Relative Defect Introduction

Incident particles, depending on their mass and energy, can trigger a vastly different recoil spectrum, as shown in Fig. 4. An obvious question is whether this affects the relative introduction probability of different defects. To address this topic further, a DDD of $2 \times 10^{11}$ MeV g$^{-1}$ is chosen. With the help of a NIEL with 21 eV, the required fluence of 1 and 3 MeV electrons as well as 1 MeV protons is calculated and the associated number of defects introduced is extracted from Table II with the help of a linear fit. The use of a $E_\mathrm{d} = 21$ eV is essential for this comparison, since only this threshold energy correctly quantifies the relative contribution of the H1-H4 defects in the overall displacement damage, as previously discussed. The absolute and relative defect concentrations are reported in the pie charts in Fig. 7.

The recoil spectrum triggered by the particles has a significant role in defining the relative concentration of the individual defects. As the energetic content of the recoil spectrum increases, it is more likely that defects with high $E_\mathrm{d}$ – such as H2 and H4 – are generated, and defects with lower $E_\mathrm{d}$ – such as H1 and H3 – account for a smaller fraction of the total damage.

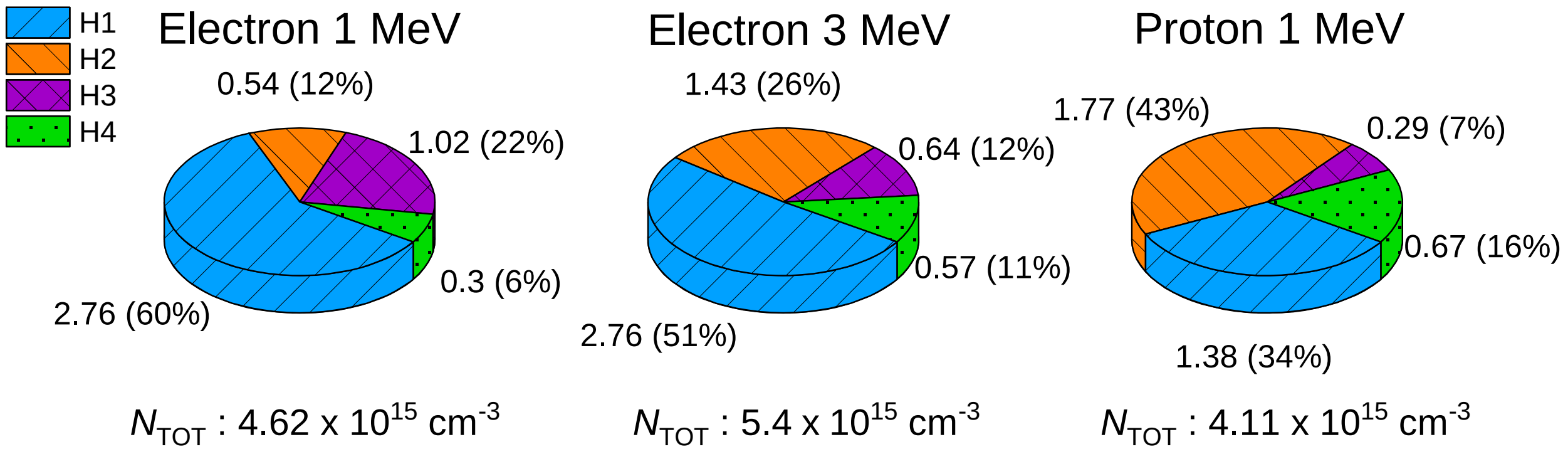

FIG. 7: Absolute (in units of $10^{15}$ cm$^{-3}$) and relative ($N_{t_0}/N_{TOT}$) defect concentration for the same amount of DDD = $2 \times 10^{11}$ MeV g$^{-1}$ introduced by 1 and 3 MeV electrons and 1 MeV protons in p-type GaAs component cells, computed with $E_d = 21$ eV. The total sum of the defects $N_{TOT}$ is also reported for each case.

### C. Macroscopic Parameter Degradation

Due to the different electronic parameters found in H1-H4, a different damaging effectiveness with respect to the solar cell macroscopic electrical parameters is expected. The ultimate cause of a solar cell electrical parameter degradation is the reduction of the carrier lifetime $\tau_{n,p}$ within the solar cell neutral ($J_{SC}$) and depleted regions ($V_{OC}$ and $J_{02}$), which is related to the defect parameters by $\tau_{n,p}^{-1} \propto \sigma_{n,p} v_{th} N_{t_0}$. If the electronic structure of the defect does not depend on the type and energy of the irradiating particle, its capture properties $\sigma_{n,p}$ are well defined and identical in the different irradiated cells. As a result, the degradation due to this class of defects occurs due to an increase of $N_{t_0}$ as more displacements occur within the semiconductor. The damaging effectiveness of each defect on the solar cell $J_{SC}$, $J_{02}$ and $V_{OC}$ can thus be estimated by means of correlation coefficients $r$ resulting from the Pearson analysis between the parameter remaining factors and the defect $N_{t_0}$. When $r$ approaches +1 (-1), a significant positive (negative) correlation between the parameters is found. The Pearson's $r$ coefficients for H1-H4 defects are reported in the color map of Fig. 8. Moreover, the Pearson analysis is extended to the DDD computed with $E_d = 21$ eV ($D_{21eV}$), which is representative of the actual scaling of the relative abundance of the defects H1-H4 with particle energy, as discussed earlier.

All defects exhibit a negative correlation with $J_{SC}$ and therefore, they are all damaging $J_{SC}$ to a certain extent. The good correlation of the $J_{SC}$ degradation with $D_{21eV}$ ($r = -0.98$) indicates that it is likely that H1-H4 are the main defects responsible for the $J_{SC}$ damage. These results are fully in line with the $J_{SC}$ fit in Fig. 1. The total number of defects H1-H4 introduced by particles of different energies varies by a factor of 1.3 at most for the same amount of $D_{21eV}$, according to the findings shown in Fig. 7. Thus the good correlation of $J_{SC}$ with $D_{21eV}$ can also be interpreted in terms of a good correlation to the sum of the defects. From this it can be concluded that the minority carrier capture cross sections of H1-H4 are in a similar range and cannot differ by orders of magnitude.

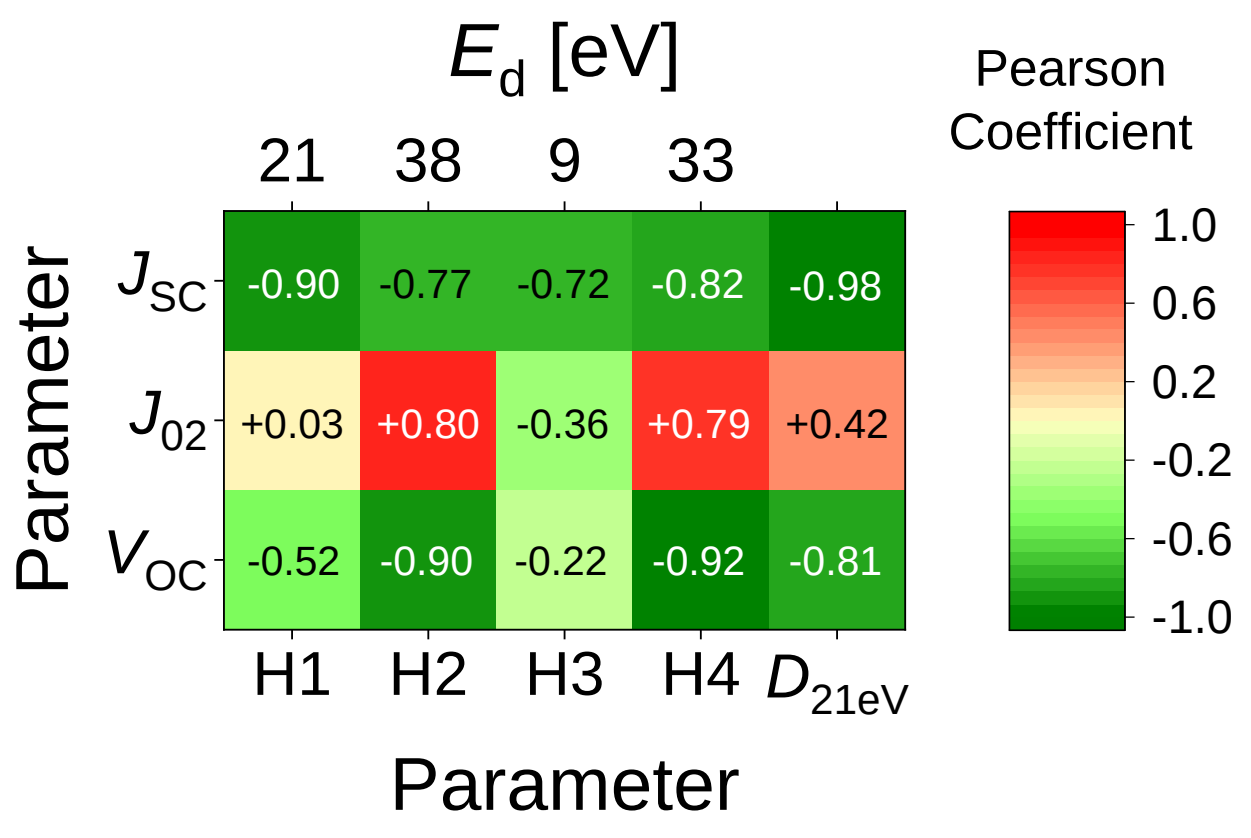

FIG. 8: Pearson's $r$ correlation coefficients of the H1, H2, H3, H4 defect concentration and the DDD computed with $E_d = 21$ eV ($D_{21eV}$) over the solar cell $J_{SC}$, $J_{02}$ and $V_{OC}$ remaining factors. A value of -1 expresses a good correlation for $J_{SC}$ and $V_{OC}$, whereas +1 expresses a good correlation for $J_{02}$.

On the other hand, $V_{OC}$ and $J_{02}$ are mostly damaged by the recombination centers H2 and H4. The defect H1 has a negligible effect on $J_{02}$, which can be attributed to its asymmetrical position within the band gap, combined with a low $\sigma_p$. Despite its apparent deep position in the band gap, the defect H3 does not affect $V_{OC}$ and $J_{02}$. Hence, it is possible that the defect presents a strong capture asymmetry resulting in a low minority-carrier capture cross section, or most likely, the energy level extracted from the trap signature is overestimated by the substantial effect of a temperature-dependent $\sigma_p(T)$. The latter theory is supported by experimental observations on some deep levels in III-V materials[30], and also by some capture theories[31].

According to Fig. 7, particles with higher relativistic mass are expected to damage $V_{OC}$ and $J_{02}$ more for a similar amount of 21 eV NIEL introduced in the cell, because these create a higher contribution of H2 and H4 defects. The best fit of the electron $J_{02}$ data with $E_d = 35$ eV, as shown in Fig. 1, is thus in line with the $E_d$ values found for H2 and H4. These defects cannot be formed by 0.5 MeV electrons, whose maximum energy transferable to a Ga or As atom is $\sim 23$ eV. The only defects that a 0.5 MeV electron can introduce in GaAs cells are H1 and H3, which exhibit a very low correlation coefficient with $J_{02}$ according to the Pearson map in Fig. 8. The $J_{02}$ degradation observed for 0.5 MeV electrons, shown in Fig. 1 (C), can be only reconciled by postulating another recombination center characterized by $E_{\mathrm{d}} < 23$ eV. This defect may be energetically located in the upper half of the band gap, out of the detectable range of admittance spectroscopy.

### D. Proton-Electron Equivalence Factors

Although the Pearson analysis yields a good correlation with $r = 0.8$ and 0.79 between the degradation of $J_{02}$ and the concentration of the H2 and H4 defects, when these mid-gap defects are introduced by protons, the damage to $J_{02}$ is more pronounced compared to the equivalent electron case, as shown in Fig. 1 (C) and (D). A plausible explanation may be the presence of defects out of the detectable range with a different introduction rate for protons and electrons. Another possible reason could be ascribed to a different capture capability of the same mid-gap defects introduced by protons and electrons. The latter explanation can be rationalized to a certain degree with the help of the admittance spectroscopy results. Analyzing the H2 defect, the distribution width $\lambda^{\mathrm{H2}}$ is found to be not related to the DDD, but rather to the distribution of recoil energies. Using an 80% cutoff in the relative NIEL contribution plot in Fig. 4, the energy $E_{80\%}$ is determined. It represents the recoil energy up to which 80% of the total displacement damage of the incident particle is deposited. As illustrated in Fig. 9 (A), a clear correlation is found between $E_{80\%}$ and $\lambda^{\mathrm{H2}}$.

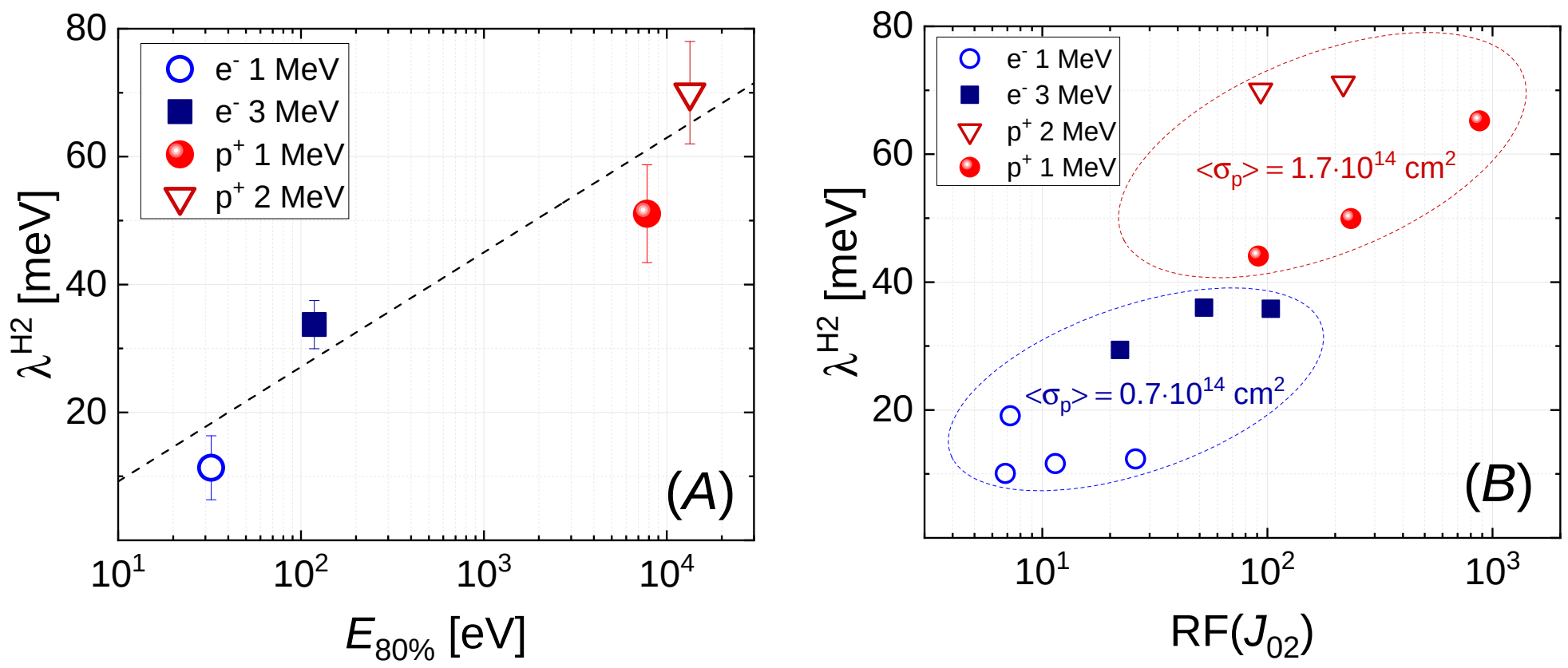

FIG. 9: (A) H2 distribution width $\lambda^{\mathrm{H2}}$ as a function of the recoil spectrum $E_{80\%}$ and (B) $\lambda^{\mathrm{H2}}$ as a function of the $J_{02}$ remaining factor in irradiated GaAs component cells. The average capture cross-section $\langle\sigma_{\mathrm{p}}\rangle$ for the proton- and electron-induced H2 defect is also reported.

This is rationalized considering that higher $E_{80\%}$ values imply larger clusters of defects. The higher electronic disorder of these regions leads to a wide distribution of available energy levels within the band gap, as already observed in literature via DLTS in GaAs cells irradiated with neutrons[32] and protons[33]. Irradiated cells with higher $\lambda^{\mathrm{H2}}$ also exhibit a higher $J_{02}$ remaining factor, as shown in Fig. 9 (B). This suggests that the larger band-tails around H2 may be linked to an increased capture capability of the defect in the SCR, thus affecting its damaging effectiveness with respect to $J_{02}$. In the limit of the uncertainty associated with the fitting of the admittance data, a higher $\sigma_{\mathrm{p}}$ is also observed for proton-irradiated cells compared to the electron case, which causes a slight blueshift of the H2 peak in the $-f\mathrm{d}C/\mathrm{d}f$ plot of Fig. 3 (A). This shows how a different particle recoil spectrum can impact the damaging effectiveness of a defect with respect to $J_{02}$. Although it might not be the only reason, this phenomenon can be linked to the existence of the electrical $\mathrm{R_{ep}}$ factor found for $J_{02}$. In the $V_{\mathrm{OC}}$ case, the existence of the $\mathrm{R_{ep}}$ factor is a direct consequence of the fact that for a moderately high dose level, the $V_{\mathrm{OC}}$ is dominated by $J_{02}$.

## VIII. CONCLUSION

In conclusion, temperature-dependent admittance spectroscopy was applied to the study of proton and electron radiation damage in GaAs component solar cells. With the help of a dedicated fitting routine, four majority carrier defects labeled H1-H4 were identified. The key benefit of admittance spectroscopy in contrast to analyzing macroscopic cell performance parameters is the fact that the radiation effects can be analyzed one step closer to the actual physical defect processes. The results depend to a lesser degree on the detailed electrical device design and thus can be interpreted more easily in terms of material-specific properties.

In particular the combination of admittance spectroscopy with the NIEL fitting of the defect introduction rate was shown to result in a unique threshold energy for each of the defects, which provides additional insight into their crystallographic configuration. Moreover, it was established that a NIEL with a threshold energy of 21 eV accurately describes the scaling of the relative abundance of H1-H4 with the energy of the incident particle. For a given DDD, the relative fraction of high-energy defect increases with the relativistic mass of the incident particle.

A Pearson analysis was then used to link these results on a defect level to the degradation of the macroscopic cell parameters. The $J_{\mathrm{SC}}$ degradation shows a direct correlation with the combined H1-H4 DDD, analytically described by a threshold energy of 21 eV, from which a similar dam-aging effectiveness of each individual H1-H4 defect on $J_{SC}$ can be concluded. For $J_{02}$, and thus also for $V_{OC}$ at high doses, the energy level of the defect in the band gap is important. Therefore, H2 and H4 play a major role for these parameters. This explains the fact that a threshold energy of 35 eV describes the DDD dependence of $J_{02}$ very well, except for low-energy electrons. In addition, H2 was found to be more damaging if introduced by protons than by electrons. This behavior was traced back to a broader distribution of the H2 density of states in the band gap in the proton case. It is in line with the existence of an $R_{ep}$ factor for $J_{02}$ and $V_{OC}$ that corrects for the less damaging nature of electron DDD as compared to proton DDD.

## ACKNOWLEDGMENTS

The authors would like to thank Dr. B. Boizot, Dr. S. Park and S. Taylor for providing access to useful samples; H. Janker, M. Nikusch, H. Hofer, and Dr. A. Gerhard for providing advice in processing the samples; M. Wildfeuer for fruitful discussions.

## DATA AVAILABILITY

The data that supports the findings of this study are available within the article.

## REFERENCES

[1] S. R. Messenger, G. P. Summers, E. A. Burke, R. J. Walters, and M. A. Xapsos, "Modeling solar cell degradation in space: A comparison of the NRL displacement damage dose and the JPL equivalent fluence approaches," Progress in Photovoltaics: Research and Applications **9**, 103–121 (2001).

[2] A. Pillukat, K. Karsten, and P. Ehrhart, "Point defects and their reactions in e -irradiated GaAs investigated by x-ray-diffraction methods," Physical Review B **53**, 7823–7835 (1996).

[3] N. Chen, S. Gray, E. Hernandez-Rivera, D. Huang, P. D. LeVan, and F. Gao, "Computational simulation of threshold displacement energies of GaAs," Journal of Materials Research **32**, 1555–1562 (2017).

[4] ESA/ESTEC, "ECSS-E-ST-20-08C Space Engineering - Photovoltaics assemblies and components," , 1–197 (2012).

[5] C. Baur, M. Gervasi, P. Nieminen, S. Pensotti, P. Rancoita, and M. Tacconi, "NIEL DOSE DEPENDENCE FOR SOLAR CELLS IRRADIATED WITH ELECTRONS AND PROTONS," in *Astroparticle, Particle, Space Physics and Detectors for Physics Applications* (WORLD SCIENTIFIC, 2014) pp. 692–707.

[6] C. Pellegrino, A. Gagliardi, and C. G. Zimmermann, "Difference in space-charge recombination of proton and electron irradiated GaAs solar cells," Progress in Photovoltaics: Research and Applications **27**, 379–390 (2019).

[7] R. Campesato, C. Baur, M. Casale, M. Gervasi, E. Gombia, E. Greco, A. Kingma, P. G. Rancoita, D. Rozza, and M. Tacconi, "NIEL DOSE and DLTS Analyses on Triple and Single Junction solar cells irradiated with electrons and protons," in *2018 IEEE 7th World Conference on Photovoltaic Energy Conversion (WCPEC) (A Joint Conference of 45th IEEE PVSC, 28th PVSEC & 34th EU PVSEC)* (IEEE, 2018) pp. 3768–3772, arXiv:1811.11583.

[8] T. P. Weiss, A. Redinger, D. Regesch, M. Mousel, and S. Siebentritt, "Direct Evaluation of Defect Distributions From Admittance Spectroscopy," IEEE Journal of Photovoltaics **4**, 1665–1670 (2014).

[9] A. I. Baranov, A. S. Gudovskikh, D. A. Kudryashov, A. A. Lazarenko, I. A. Morozov, A. M. Mozharov, E. V. Nikitina, E. V. Pirogov, M. S. Sobolev, K. S. Zelentsov, A. Y. Egorov, A. Darga, S. Le Gall, and J.-P. Kleider, "Defect properties of InGaAsN layers grown as sub-monolayer digital alloys by molecular beam epitaxy," Journal of Applied Physics **123**, 161418 (2018).

[10] M. Levinshtein, S. Rumyantsev, and M. Shur, *Handb. Ser. Semicond. Parameters. Vol. 2* (1996).

[11] P. Blood and J. W. Orton, *The electrical characterization of semiconductors: majority carriers and electron states*, 1st ed. (Academic Press, 1992, New York, 1990).

[12] R. Hoheisel and A. W. Bett, "Experimental Analysis of Majority Carrier Transport Processes at Heterointerfaces in Photovoltaic Devices," IEEE Journal of Photovoltaics **2**, 398–402 (2012).

[13] M. Burgelman, P. Nollet, and S. Degrave, "Modelling polycrystalline semiconductor solar cells," Thin Solid Films **361-362**, 527–532 (2000).

[14] W. Hu, J. Xie, H. W. Chau, and B. C. Si, "Evaluation of parameter uncertainties in nonlinear regression using Microsoft Excel Spreadsheet," Environmental Systems Research **4**, 4 (2015).

[15] Insoo Jun, "Effects of secondary particles on the total dose and the displacement damage in space proton environments," IEEE Transactions on Nuclear Science **48**, 162–175 (2001).

[16] J. Lindhard, V. Nielsen, M. Scharff, and P. V. Thomsen, "Integral equations governing radiation effects," Mat. Fys. Medd. Dan. Vid. Selsk (1963).

[17] M. J. Boschini, C. Consolandi, M. Gervasi, S. Giani, D. Grandi, V. Ivanchenko, P. Nieminem, S. Pensotti, P. G. Rancoita, and M. Tacconi, "NUCLEAR AND NON-IONIZING ENERGY-LOSS OF ELECTRONS WITH LOW AND RELATIVISTIC ENERGIES IN MATERIALS AND SPACE ENVIRONMENT," (2012) pp. 961–982, arXiv:1111.4042.

[18] M. J. Boschini, C. Consolandi, M. Gervasi, S. Giani, D. Grandi, V. Ivanchenko, P. Nieminem, S. Pensotti, P. G. Rancoita, and M. Tacconi, "An expression for the Mott cross section of electrons and positrons on nuclei with Z up to 118," Radiation Physics and Chemistry **90**, 39–66 (2013), arXiv:1304.5871.

[19] M. Boschini, C. Consolandi, M. Gervasi, S. Giani, D. Grandi, V. Ivanchenko, S. Pensotti, P. Rancoita, and M. Tacconi, "Nuclear and Non-Ionizing Energy-Loss for Coulomb Scattered Particles from Low Energy up to Relativistic Regime in Space Radiation Environment," in *Cosmic Rays for Particle and Astroparticle Physics* (WORLD SCIENTIFIC, 2011) pp. 9–23, arXiv:1011.4822.

[20] M. Boschini, P. Rancoita, and M. Tacconi, "SR-NIEL Calculator: Screened Relativistic (SR) Treatment for Calculating the Displacement Damage and Nuclear Stopping Powers for Electrons, Protons, Light- and Heavy- Ions in Materials," [Online] available at INFN sez. Milano-Bicocca, Italy [2020,07]: http://www.sr-niel.org/. (2020).

[21] F. Gao, N. Chen, E. Hernandez-Rivera, D. Huang, and P. D. LeVan, "Displacement damage and predicted non-ionizing energy loss in GaAs," Journal of Applied Physics **121**, 095104 (2017).

[22] J. F. Ziegler, M. D. Ziegler, and J. P. Biersack, "SRIM - The stopping and range of ions in matter (2010)," Nuclear Instruments and Methods in Physics Research, Section B: Beam Interactions with Materials and Atoms **268**, 1818–1823 (2010).

[23] T. Mattila and R. M. Nieminen, "Direct Antisite Formation in Electron Irradiation of GaAs," Physical Review Letters **74**, 2721–2724 (1995).

[24] J. C. Bourgoin, H. J. von Bardeleben, and D. Stiévenard, "Native defects in gallium arsenide," Journal of Applied Physics **64**, R65–R92 (1988).

[25] J. C. Bourgoin and M. Zazoui, "Irradiation-induced degradation in solar cell: characterization of recombination centres," Semiconductor Science and Technology **17**, 453–460 (2002).

[26] H. J. von Bardeleben, D. Stiévenard, D. Deresmes, A. Huber, and J. C. Bourgoin, "Identification of a defect in a semiconductor: EL 2 in GaAs," Physical Review B **34**, 7192–7202 (1986).

[27] D. Stievenard, X. Boddaert, and J. C. Bourgoin, "Irradiation-induced defects in p-type GaAs," Physical Review B **34**, 4048–4058 (1986).

[28] D. Pons and J. C. Bourgoin, "Irradiation-induced defects in GaAs," Journal of Physics C: Solid State Physics **18**, 3839–3871 (1985).

[29] A. F. Wright and N. A. Modine, "Migration processes of the As interstitial in GaAs," Journal of Applied Physics **120**, 215705 (2016).

[30] P. Blood and J. J. Harris, "Deep states in GaAs grown by molecular beam epitaxy," Journal of Applied Physics **56**, 993–1007 (1984).

[31] J. Bourgoin and M. Lannoo, *Point Defects in Semiconductors II*, Springer Series in Solid-State Sciences, Vol. 35 (Springer Berlin Heidelberg, Berlin, Heidelberg, 1983).

[32] R. M. Fleming, D. V. Lang, C. H. Seager, E. Bielejec, G. A. Patrizi, and J. M. Campbell, "Continuous distribution of defect states and band gap narrowing in neutron irradiated GaAs," Journal of Applied Physics **107**, 123710 (2010).

[33] J. H. Warner, S. R. Messenger, R. J. Walters, S. A. Ringel, and M. R. Brenner, "A deep level transient spectroscopy study of electron and proton irradiated p+n GaAs diodes," Proceedings of the European Conference on Radiation and its Effects on Components and Systems, RADECS **57**, 282–285 (2009).

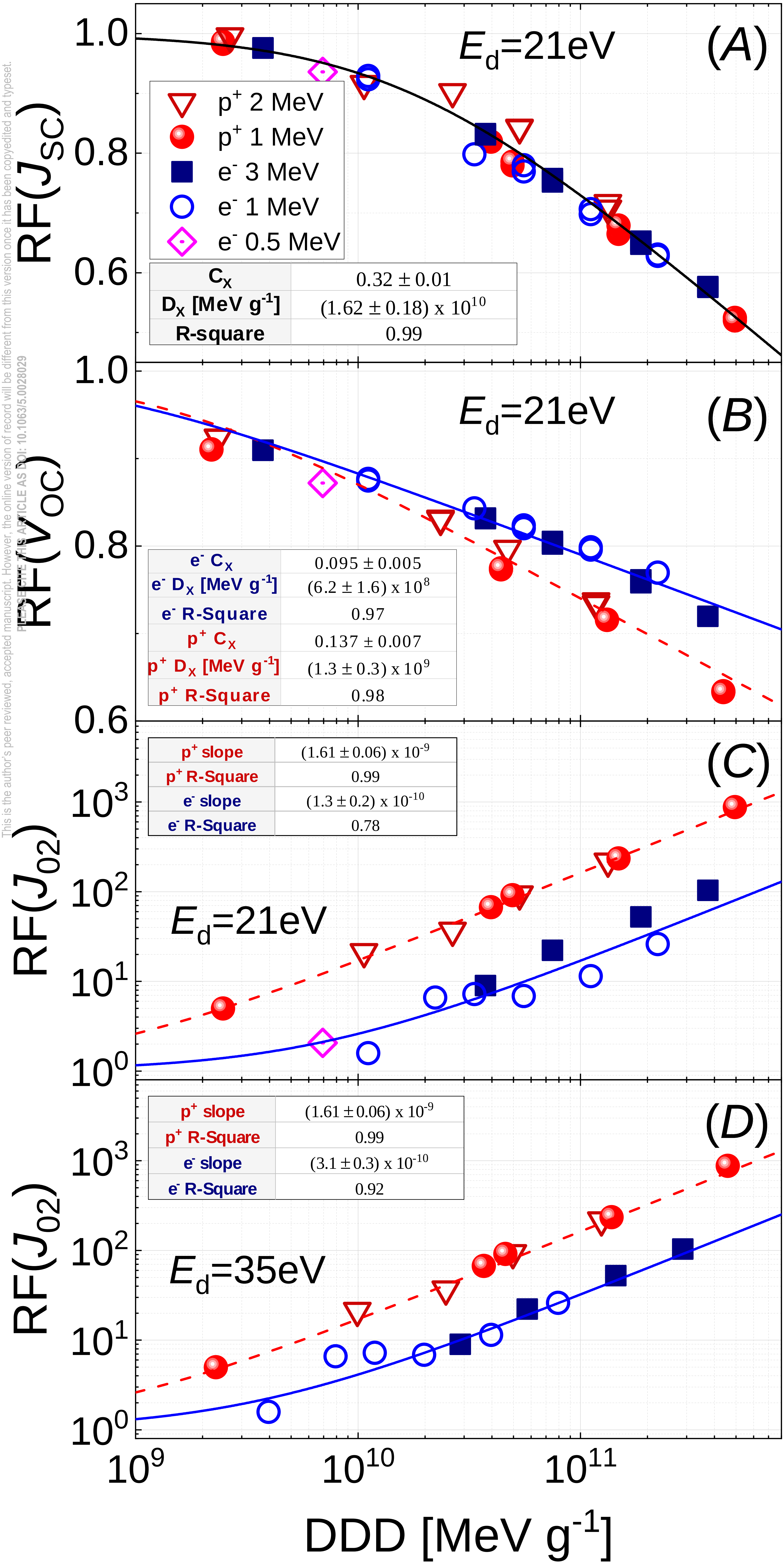
(A)
E_d=21eV
RF(J_SC)
p+ 2 MeV
p+ 1 MeV
e- 3 MeV
e- 1 MeV
e- 0.5 MeV
C_X 0.32 ± 0.01
D_X [MeV g^-1] (1.62 ± 0.18) x 10^10
R-square 0.99
(B)
E_d=21eV
RF(V_OC)
e- C_X 0.095 ± 0.005
e- D_X [MeV g^-1] (6.2 ± 1.6) x 10^8
e- R-Square 0.97
p+ C_X 0.137 ± 0.007
p+ D_X [MeV g^-1] (1.3 ± 0.3) x 10^9
p+ R-Square 0.98
(C)
E_d=21eV
RF(J_02)
p+ slope (1.61 ± 0.06) x 10^-9
p+ R-Square 0.99
e- slope (1.3 ± 0.2) x 10^-10
e- R-Square 0.78
(D)
E_d=35eV
RF(J_02)
p+ slope (1.61 ± 0.06) x 10^-9
p+ R-Square 0.99
e- slope (3.1 ± 0.3) x 10^-10
e- R-Square 0.92
DDD [MeV g^-1]

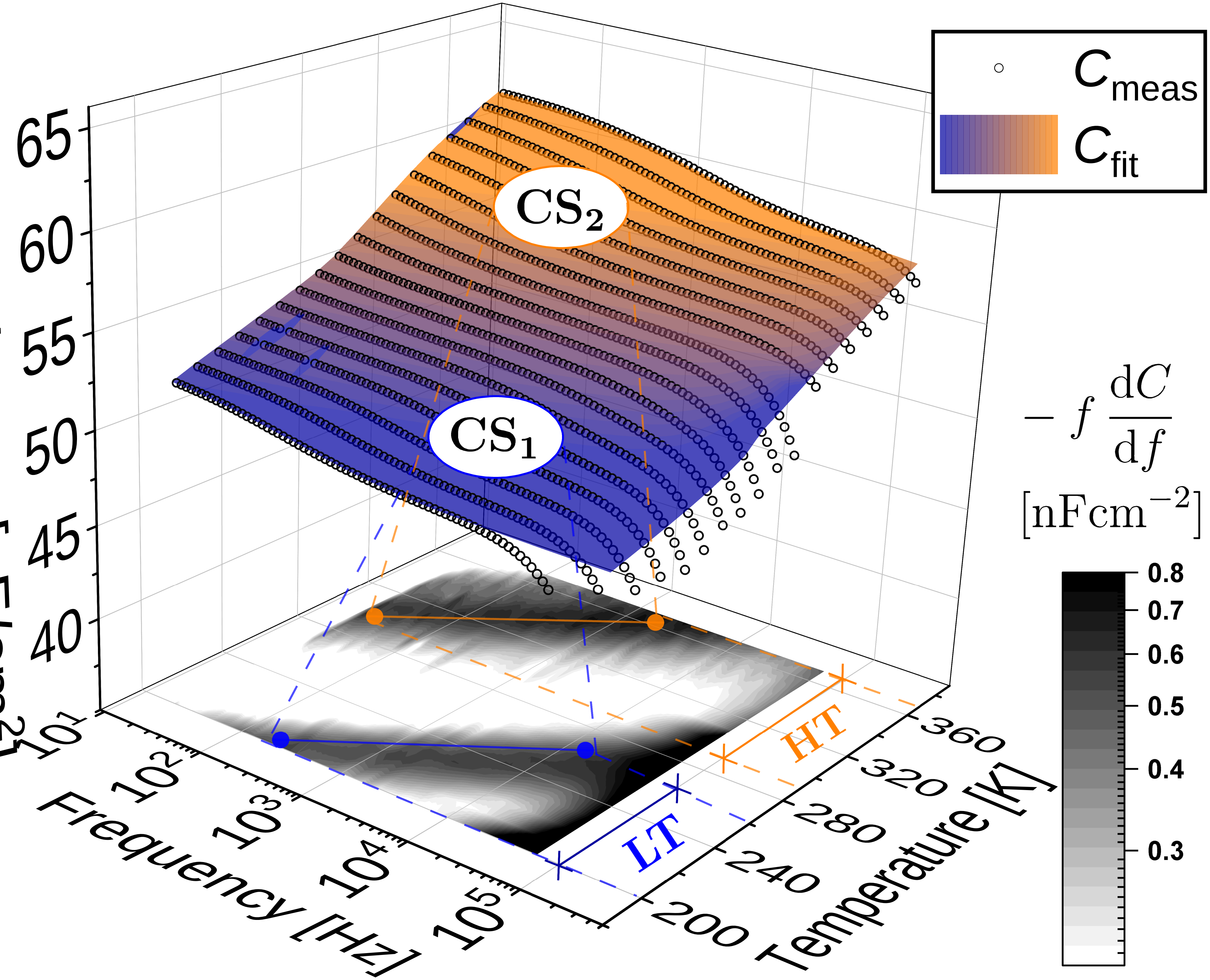

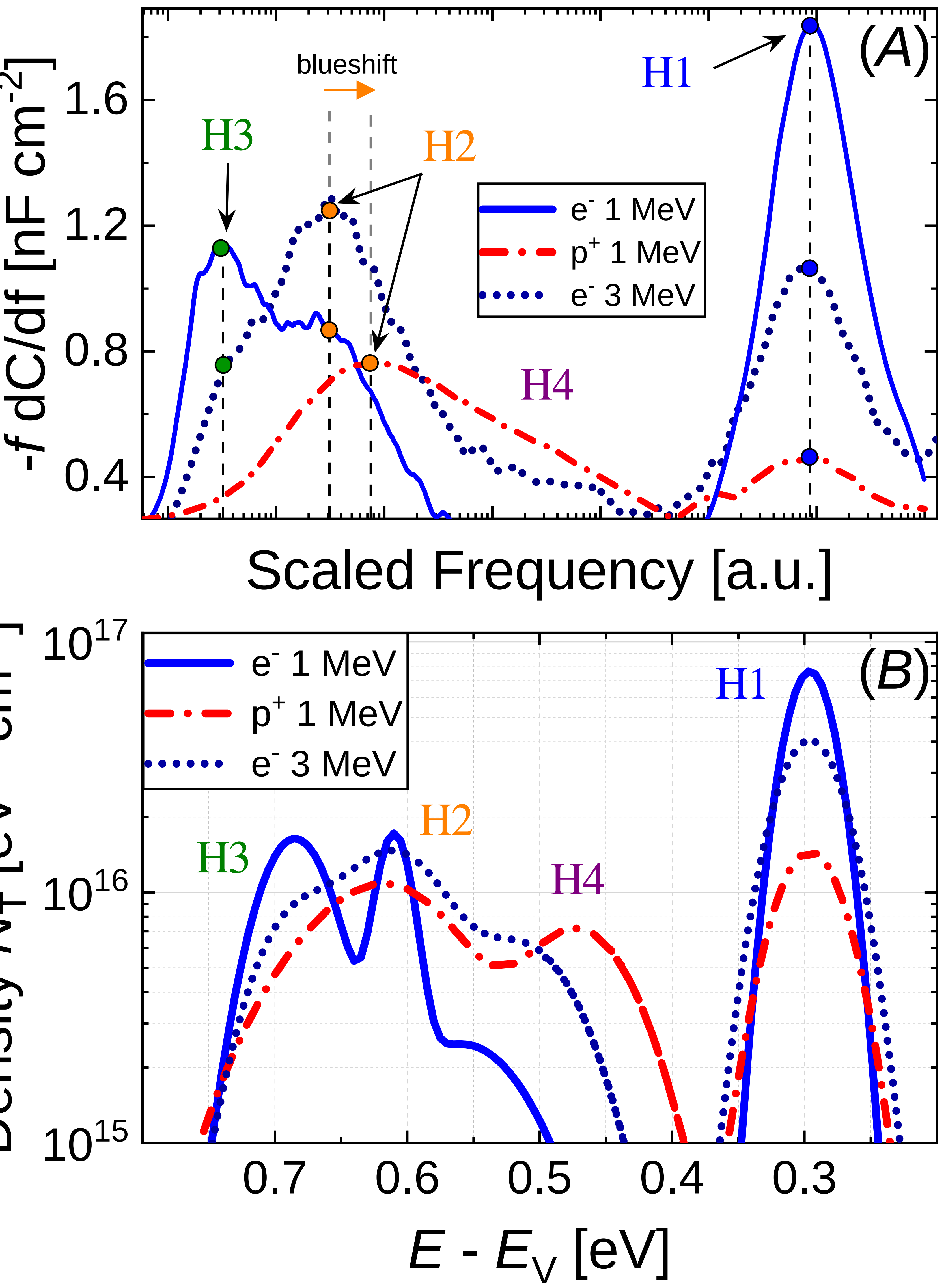

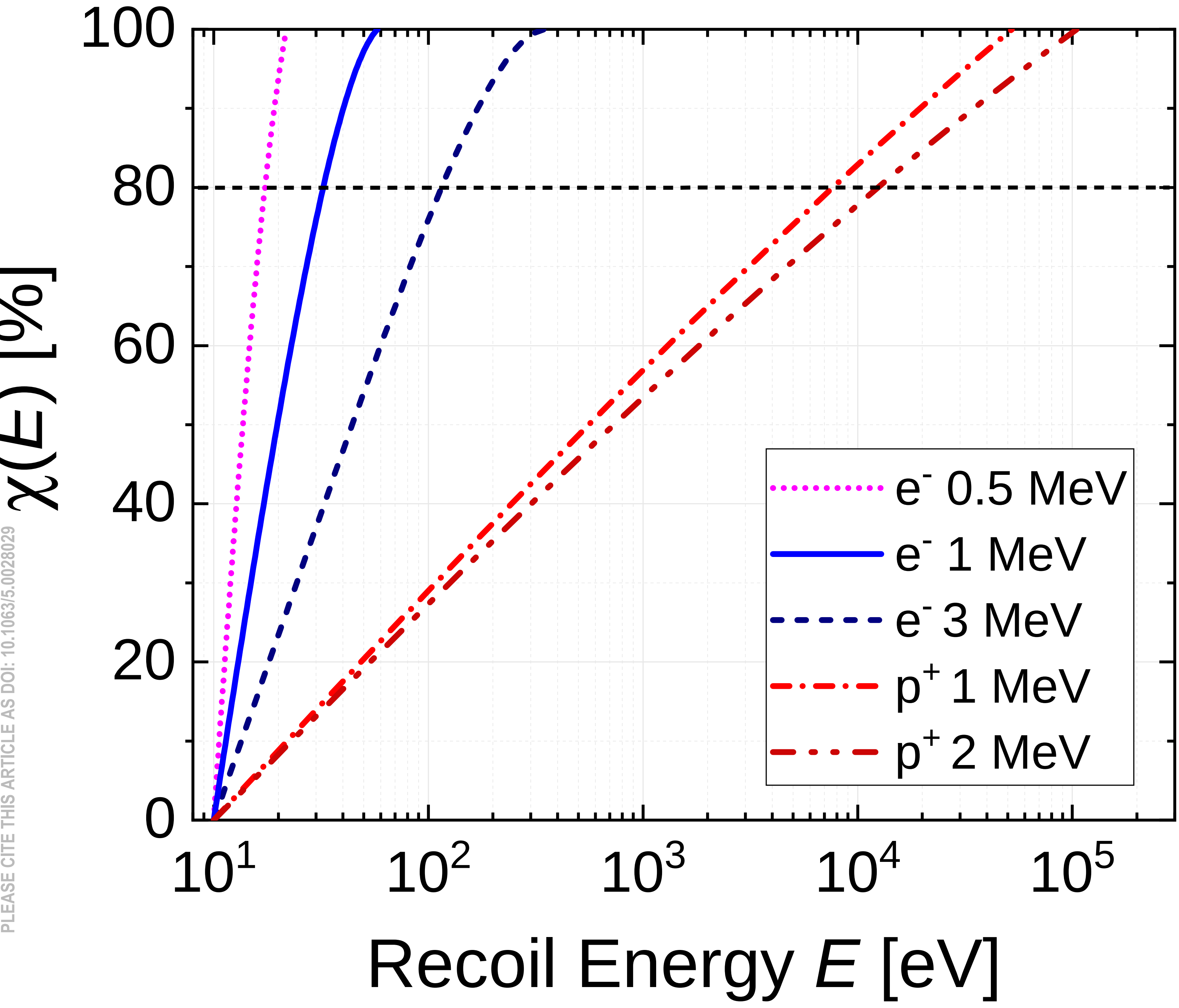

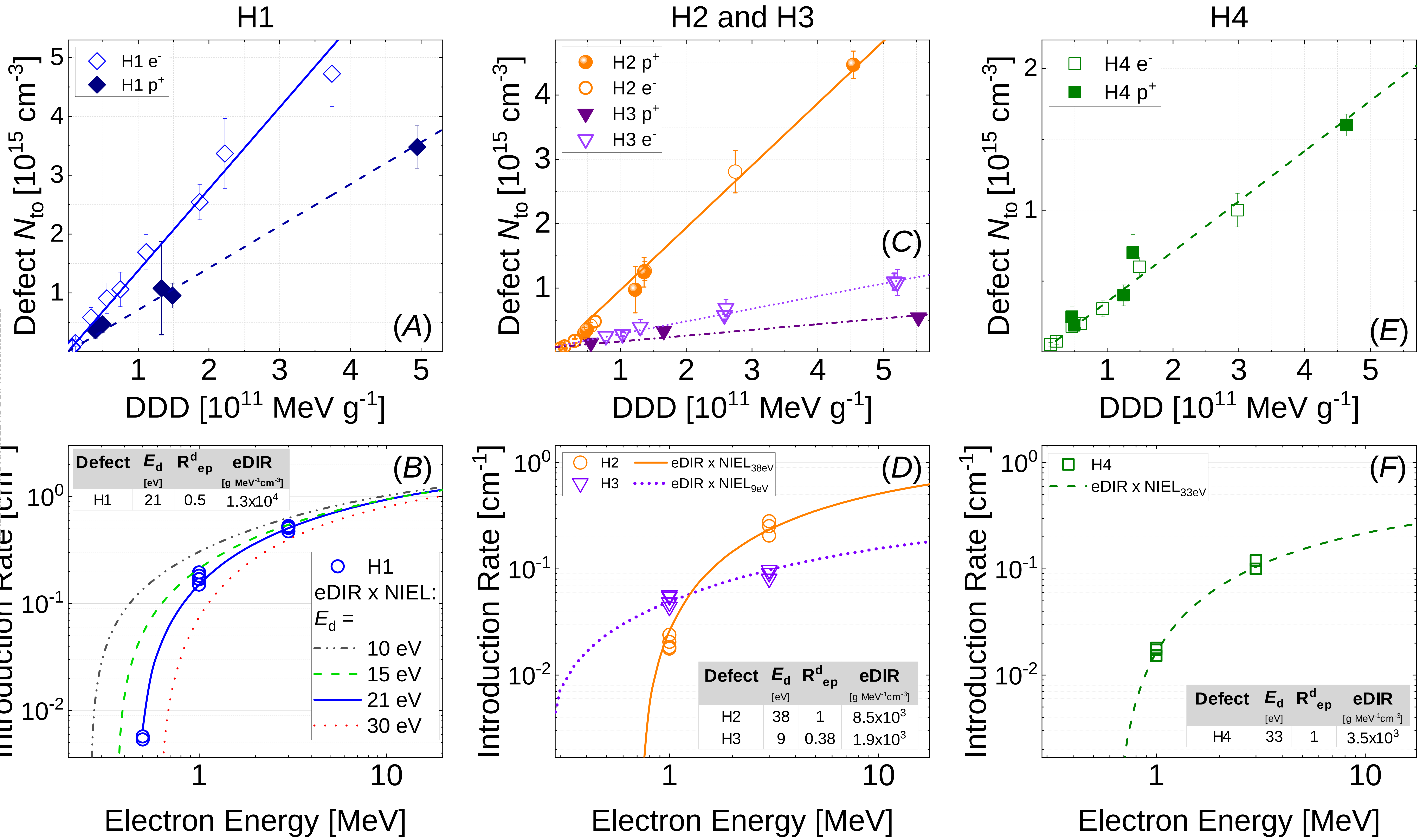
H1
H2 and H3
H4
H1 e-
H1 p+
(A)
Defect $N_{to}$ [$10^{15}$ cm$^{-3}$]
DDD [$10^{11}$ MeV g$^{-1}$]
H2 p+
H2 e-
H3 p+
H3 e-
(C)
H4 e-
H4 p+
(E)
(B)
Introduction Rate [cm$^{-1}$]
Electron Energy [MeV]
Defect | $E_d$ [eV] | $R^d_{ep}$ | eDIR [g MeV$^{-1}$cm$^{-3}$]
H1 | 21 | 0.5 | 1.3x$10^4$
H1
eDIR x NIEL:
$E_d$ =
10 eV
15 eV
21 eV
30 eV
(D)
H2
H3
eDIR x NIEL$_{38eV}$
eDIR x NIEL$_{9eV}$
Defect | $E_d$ [eV] | $R^d_{ep}$ | eDIR [g MeV$^{-1}$cm$^{-3}$]
H2 | 38 | 1 | 8.5x$10^3$
H3 | 9 | 0.38 | 1.9x$10^3$
(F)
H4
eDIR x NIEL$_{33eV}$
Defect | $E_d$ [eV] | $R^d_{ep}$ | eDIR [g MeV$^{-1}$cm$^{-3}$]
H4 | 33 | 1 | 3.5x$10^3$

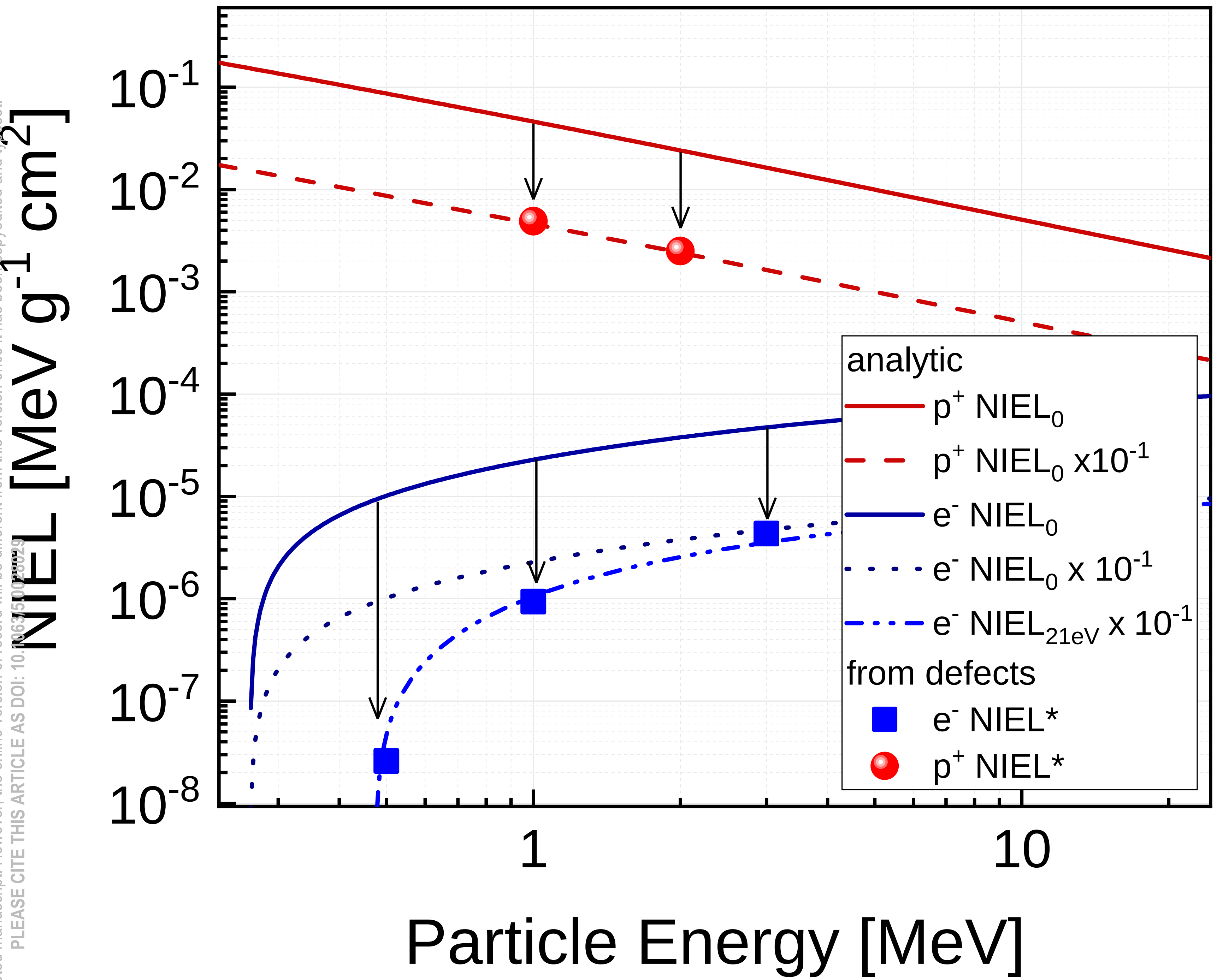

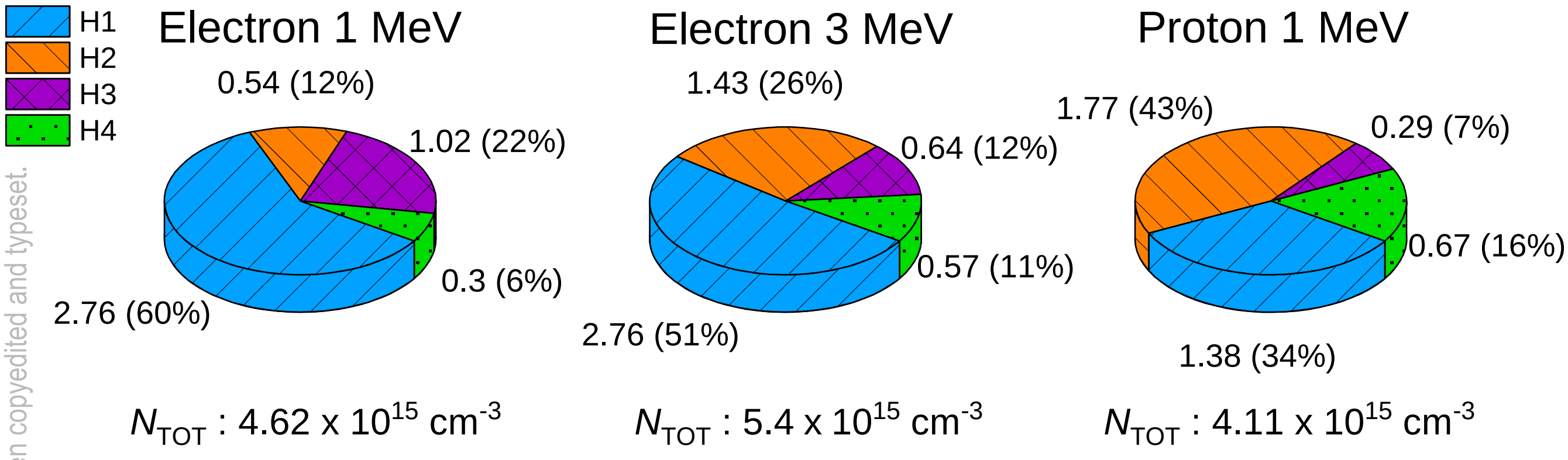

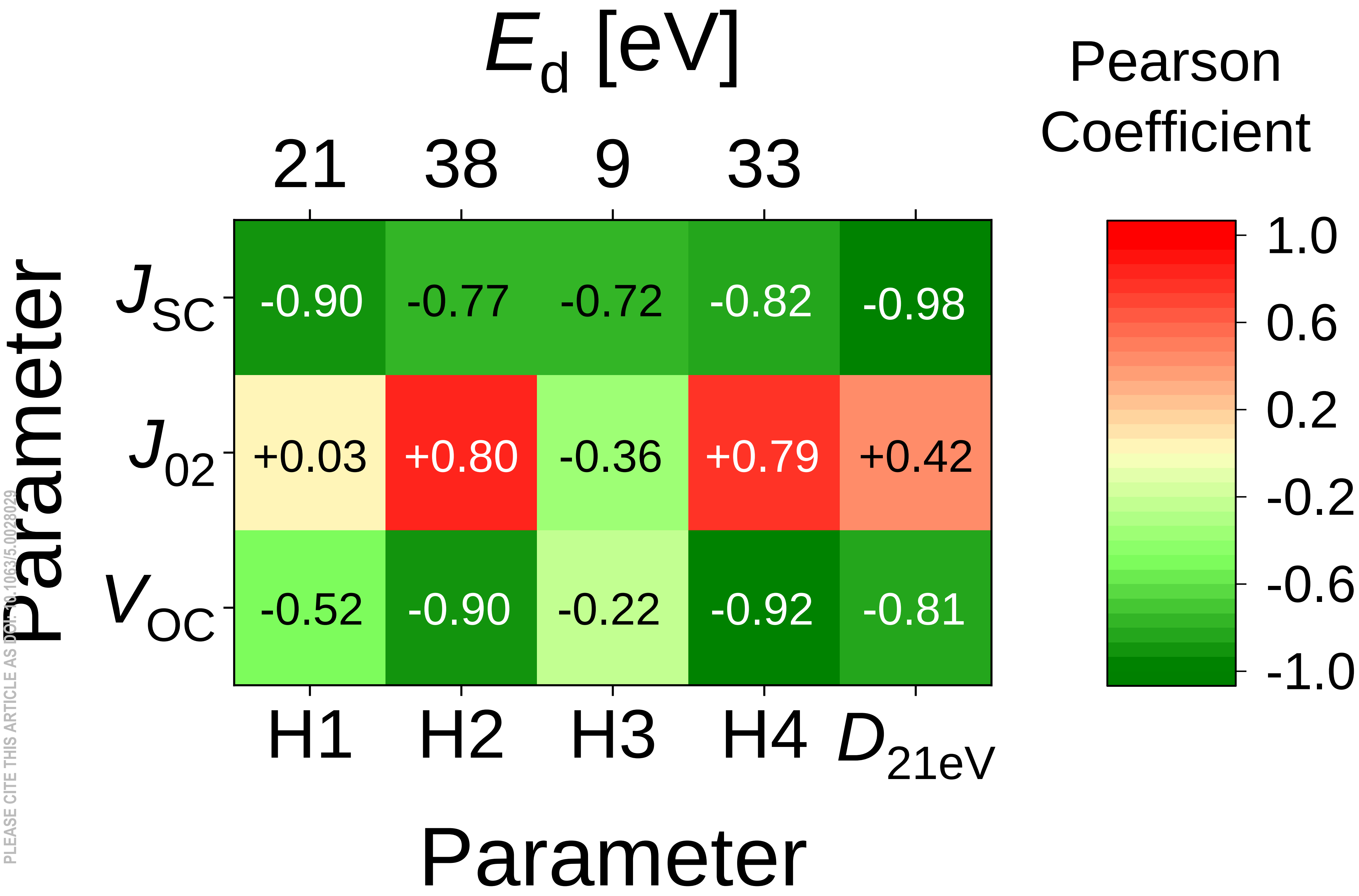

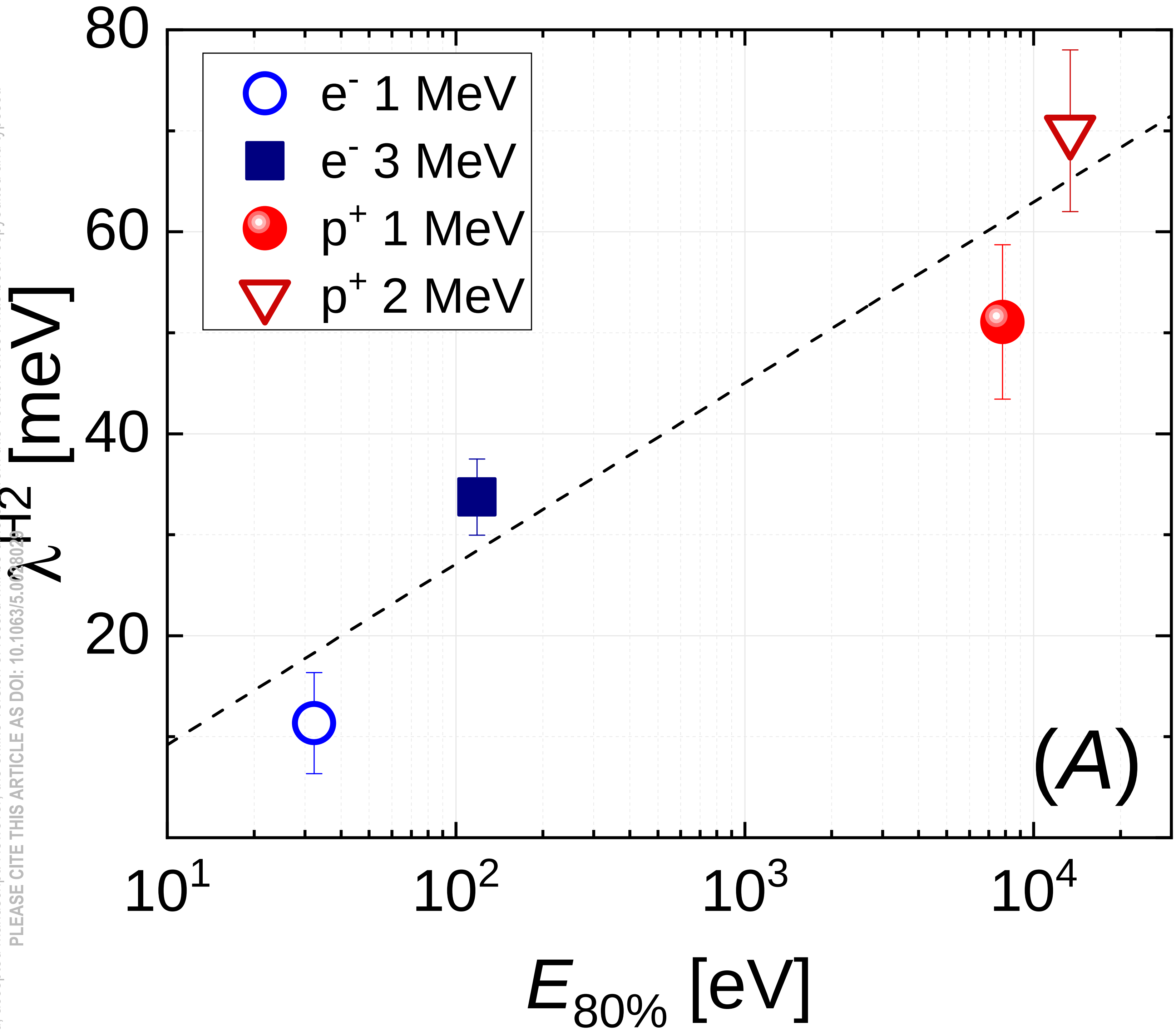

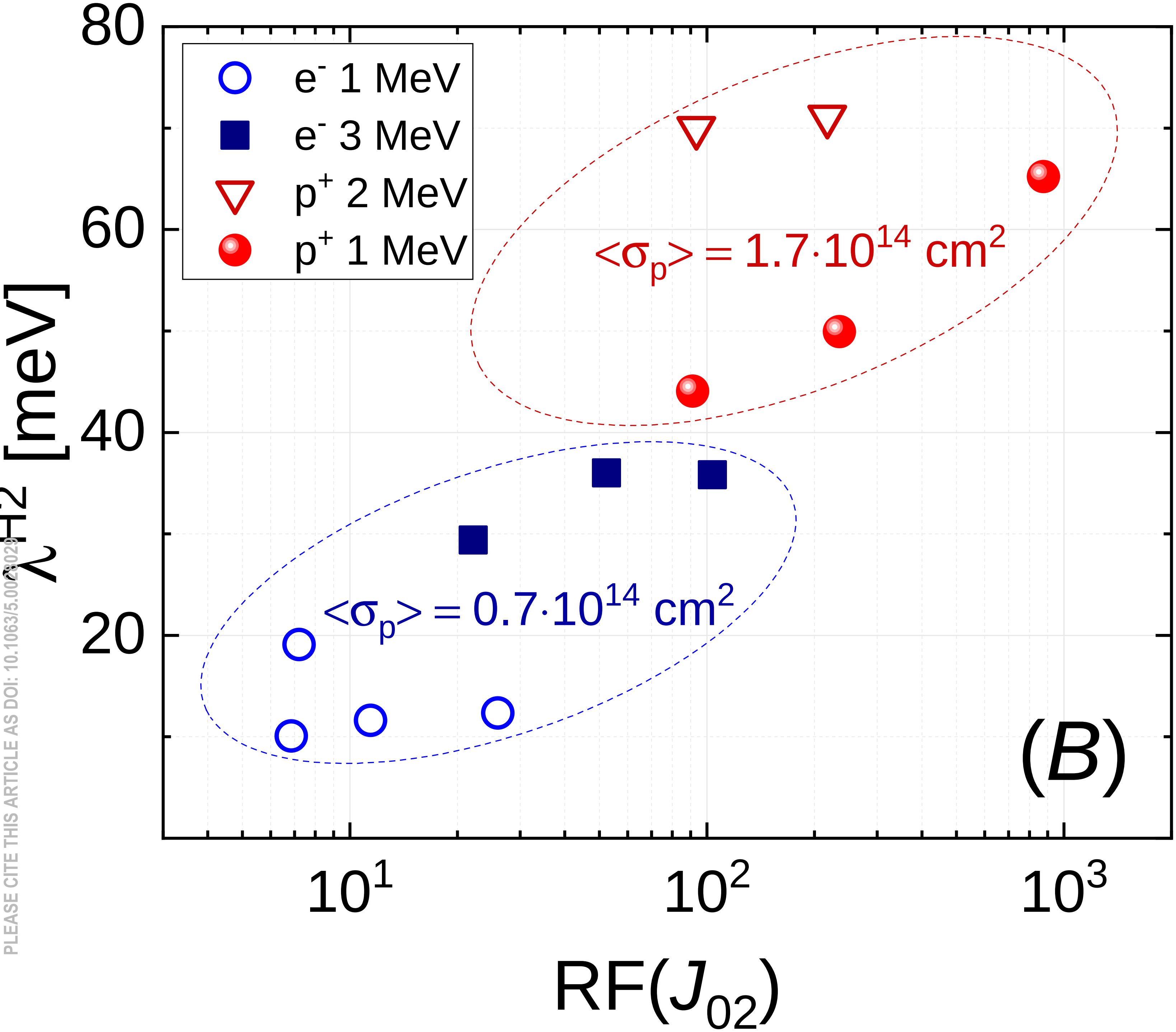